\documentstyle[prc,aps,preprint,psfig]{revtex}
\begin{document}
\draft
\title{\bf Low energy behavior of the astrophysical
S-factor in radiative captures to loosely bound final states}
\author{A.M. Mukhamedzhanov$^1$ and F.M. Nunes$^2$}
\address{1) Cyclotron Institute, Texas A\&M University,
College Station, TX, 77843}
\address{2) Universidade Fernando Pessoa, 4249-004 Porto, Portugal}
\maketitle

\begin{abstract}
The low-energy behavior of the astrophysical S-factor for $E1$ direct
radiative captures a(p,$\gamma$)b leading to loosely bound final states
(b=a+p) is investigated. We derive a first-order integral
representation
for S(E) and focus on the properties around zero energy.
We show that it is the competition between various effects, namely
the remnant Coulomb barrier, the initial and final centrifugal barriers
and the binding energy, that defines the behavior of the
$S(E\rightarrow0)$.
Contrary to previous findings, we prove that $S(E\rightarrow0)$ is not
determined by the pole corresponding to the bound state.
The derivative $S'(0)$ increases with the increase of the centrifugal
barrier,
while it decreases with the charge of the target.
For $l_{i}=l_{f}+1$ the increase of the binding
energy of the final nucleus increases the derivative $S'(0)$ while
for $l_{i}=l_{f}-1$ the opposite effect is found.
We make use of our findings to explain the low energy behavior
of the S-factors related to some notorious capture reactions:
${}^7{\rm Be}(p, \gamma){}^8{\rm B}$,
${}^{14}{\rm N}(p,\gamma){}^{15}{\rm O}$,
${}^{16}{\rm O}(p,\gamma){}^{17}{\rm F}$,
${}^{20}{\rm {Ne}}(p, \gamma){}^{21}{\rm {Na}}$ and
${}^{22}{\rm Mg}(p, \gamma){}^{23}{\rm Al}$. 
\end{abstract}



\section{Introduction}
Direct radiative capture cross sections of charged particles
a(p,$\gamma$)b
drop so sharply for the low energy region of stellar environment
that it is often impossible to measure these rates at the appropriate
energies
in the laboratory \cite{adelberger98}. Then, typically, the direct
capture cross sections are measured at higher energies followed by an
extrapolation down to zero energy. Therefore, the knowledge of the
energy behavior of the cross section as $E \to 0$ is of crucial importance.

When an incident charged particle approaches a target it should
penetrate through the Coulomb and centrifugal barriers in the initial
state. The capture to a loosely bound
state is the transition from the initial
scattering state to the tail of the bound state wave
function in the final state. This tail is affected by the Coulomb and
centrifugal barriers in the final state. Thus, the radiative
capture process is affected by four barriers, two Coulomb
and  two centrifugal barriers, implicitly present through the
four parameters $\eta_i, \eta_f, l_i, l_f$.
All barriers have a similar effect on the radiative cross
section: they decrease the probability of the reaction
at sub-barrier energies.

The cross section is usually factorized into
the Gamow penetration factor, and the astrophysical S-factor:
\begin{equation}
\sigma(E)= \frac{e^{-2\,\pi\,\eta_{i}}}{E}\,S(E).
\label{radcp1}
\end{equation}
The Gamow penetration factor is no more than the
probability
for an $s$-wave proton to penetrate through the pure Coulomb barrier
(i.e. the Coulomb potential is extended down to the origin $r=0$
assuming a pointlike target).
Naturally, this factor  defines only the gross energy behavior.
In fact, as the capture to loosely bound states
occurs under the Coulomb barrier, at a large distance from the target,
the effective penetration factor is larger than the Gamow penetration
factor in Eq. (\ref{radcp1})
\cite{rolfs}.
We call {\it remnant Coulomb barrier} the remaining energy
dependence of the astrophysical factor S(E) due to the initial
Coulomb barrier, after removing the Gamow factor.
This remnant effect behaves in an opposite way to the normal barrier
behavior, namely,  it reduces the S-factor as the energy
increases, to compensate the overestimation by the Gamow factor.

It turns out  that, depending on the system, S-factors can feature
completely opposite behaviors as one approaches zero energy.
Despite the long history of S-factor calculations and the
numerous papers published on this subject, no satisfactory physical
explanation  for the different behaviors has been presented.  
In recent papers \cite{jenn98,baye2000} some aspects of the low
energy behavior of the direct capture S-factors were investigated.
In \cite{jenn98}, the low energy rise of the S-factor was attributed
to the pole in the energy plane, located at $E=-\varepsilon$
(with $\varepsilon$ being the binding energy of the final
nucleus $b=a+p$ relative to proton threshold). In \cite{baye2000},
the integral form for the first three terms of the Taylor expansion
were derived, using a potential model.
In either work, the physical reasons for specific patterns in the
S-factor
energy behavior were not considered.

If the rise of $S(E\rightarrow0)$ were in fact caused by the pole
how could one understand the decreasing $S(E\rightarrow0)$ for
${}^{16}{\rm O}(p, \gamma){}^{17}{\rm F}$(gr. st.), a case where
the pole is extremely close to the threshold?
Why should the S-factor for
${}^{16}{\rm O}(p, \gamma){}^{17}{\rm F}(2s_{1/2})$ have a steeper
slope
near the threshold than the S-factor for
${}^{7}{\rm Be}(p, \gamma){}^{8}{\rm B}$? 
What are the crucial parameters defining the behavior of the
S-factor for the direct captures to loosely bound states?
Is it possible to define general thumb rules?
It is clear that accurate numerical values for direct capture S-factors
are trivially calculated with present day technology. For that reason
analytical works have not been highly appreciated.
Regardless, this work is a search for the understanding of the general 
physical
features, based on analytical considerations.

The vast majority of the direct proton radiative capture reactions
with interest to Astrophysics, are dominated by $E1$ transitions.
We will thus concentrate on the $E1$ transition for the $a+p \to b +
\gamma$
process. We also confine ourselves to loosely bound final states.
The interest on the behavior of the S-factors for the
direct captures reactions goes beyond $p$-shell
nuclei. The heavier, $sd$-shell nuclei \cite{herndl95}, when near
the proton drip line and/or near shell closure, have very small
$Q$-values 
for the radiative capture. Thus the S-factor is dominated by single
resonant and direct captures. Consequently, the findings of this
work have implications for a great number of today's interesting
astrophysical cases.

In this paper we use an analytical/integral representation of the
transition
matrix element for the E1 process and impose the asymptotic
approximation.
Following, we identify the physical ingredients that affect the
low-energy behavior of capture reactions  to loosely bound states.
In Section \ref{sect_1} the effect of the pole of $S(E)$ at 
negative energy is specifically addressed.
The threshold behavior is studied in Section \ref{thresh}.
The competition between the remnant initial Coulomb barrier, 
the final Coulomb, the initial and final centrifugal barriers,
as well as the trace of the singularity at
$E=-\varepsilon$  are detailed and discussed. 
The S-factor dependence on the target charge is
also studied. Section \ref{application} consists of an application
to some important astrophysical capture processes.
Finally, in section \ref{summary}, a summary of the results is
presented.


\section{The S-factor around the pole}
\label{sect_1}

For E1 captures to loosely bound states, 
the reactions under consideration become purely peripheral
and a simple solid sphere model is adequate.
Such an approach is also used when calculating the non-resonant capture
in the R-matrix method \cite{barker}.
The S-factor for the capture from an initial state of $a+p$ with
orbital angular momentum $l_{i}$,
to a final state $J_f$ of $b=a+p$,
with the angular orbital momentum $l_{f}$, can be written
as \cite{barker}:
\begin{eqnarray}
S(E)&=&
{\cal A} \; \frac{e^{2\,\pi\,\eta_{i}}}{\mu_{ap}E}\, 
(k_{\gamma}\,r_{0})^{2L+1}\,
P_{l_{i}}\,                              
\nonumber\\
&\times & |\frac{1}{r_{0}^{L+1}}\, \int\limits_{r_{0}}^{\infty}
{\rm d}r\,r^{L}\,W_{-\eta_{f},l_{f}+1/2}(2\,\kappa\,r)    
\lbrack F_{l_{i}}(k,r)G_{l_{i}}(k,r_{0})- F_{l_{i}}(k,r_{0})\,
G_{l_{i}}(k,r)\rbrack |^{2},
\label{sfctr1}
\end{eqnarray}
where ${\cal A}$ is a factor depending on spin-angular characteristics,
masses, charges and constants:
\begin{eqnarray}
{\cal A}&=&2\pi \cdot
10^{4}\,\frac{(2\,J_{f}+1)}{(2\,J_{a}+1)\,(2\,J_{p}+1)}\,
\frac{1}{2\,l_{f}+1}\,(\mu_{ap})^{2\,L}\left(\frac{Z_{p}}{m_{p}^{L}}+(-1)^{L}\,
\frac{Z_{a}}{m_{a}^{L}}\right)^{2}\,
\nonumber\\
&\times &
\frac{(L+1)\,(2\,L+1)}{L}\,\frac{1}{\lbrack(2\,L+1)!!\rbrack^{2}}\,
C^{2}\,(\mu_{ap} c)^2\, m\,c^{2} \,
\frac{e^{2}}{\hbar\,c}\,\lambda_{p}^{3}\;.
\label{eq:constant}
\end{eqnarray}
Here $C$ is the asymptotic normalization coefficient of the tail of the 
projection
of the bound state wave function of the final nucleus $b$ onto the 
two-body channel
$a + $p.
The Sommerfeld parameters for initial and final
states, $\eta_{i}=Z_{a}\,e^{2}\,\mu_{ap}/k$ and
$\eta_{f}= Z_{a}\,e^{2}\,\mu_{ap}/\kappa$, are defined in terms of
the initial relative momentum of $a+p$, $k=\sqrt{2\,\mu_{ap}\,E}$,
and the bound state momentum, $\kappa=\sqrt{2\,\mu_{ap}\,\varepsilon}$,
respectively. Note that the initial relative energy is E and
the energy of the final state is $-\varepsilon$, so the photon has
energy  $E+\varepsilon$, momentum
$k_{\gamma}=(E+\varepsilon)/{\hbar\,c}$
and multipolarity $L$.
The penetration factor is defined in terms of the regular and
irregular Coulomb functions at $r=r_0$:
\begin{equation}
P_{l_{i}}=k\,r_{0}/[F_{l_{i}}^{2}(k,r_{0})+
G_{l_{i}}^{2}(k,r_{0})]  
\label{penetr1}
\end{equation}
and $W_{-\eta_{f}, l_{f}+1/2}(2\,\kappa\,r)$ is the Whittaker
function describing the radial behavior of the bound state
wave function at $r >r_{0}$, where $r_{0}$ is the solid sphere radius. 
As usual, $Z_a$ is the charge of $a$ and $\mu_{ap}$ is the reduced mass
of the
$a+p$ system. The constants used are $\lambda_{p}=0.2118$ fm for the
proton Compton
wave number and $m\,c^{2}= 931.5$ MeV for the mass atomic unit.
The factor $10^{4}$ is introduced to provide the S-factor in keV b.

When the binding energy of final state is small and the $a-p$ relative
energy is low, nuclear re-scattering $a-p$ can be neglected.
This corresponds to neglecting the term containing the irregular
Coulomb function, $G_{l_{i}}^{2}(k,r_{0})$.
One should keep in mind that whenever precise values for $S(E > 0)$ are 
needed, phase shifts should be taken into account. However, here we are 
only concerned with the qualitative behavior of S(E) around zero.
In section V comments on the accuracy of our results are presented.
Eq. (\ref{sfctr1}) then simplifies to:
\begin{equation}
S(E)= {\cal A} \; \frac{e^{2\,\pi\,\eta_{i}}}{\mu_{ap}E}\,
(k\,r_{0})\,(k_{\gamma}\,r_{0})^{2\,L+1} 
 |\frac{1}{r_{0}^{L+1}}\,\int\limits_{r_{0}}^{\infty}
{\rm d}r\,r^{L}\,W_{-\eta_{f},l_{f}+1/2}(2\,\kappa\,r)\,
F_{l_{i}}(k,r) |^{2}.
\label{sfctr2}
\end{equation}
Both, Eq. (\ref{sfctr1}) and Eq. (\ref{sfctr2}), can be used for the 
sake of our following arguments.

Next, we realize that when $k \to 0$, extending the integral in
Eq. (\ref{sfctr2})
from $r_0$ to zero is a reasonable approximation, as then the dominant
contribution comes from the asymptotic region.
Replacing the bound state wave function by the Whittaker
function down to zero radius is certainly not accurate for $k>0$,
but we have verified that the qualitative features for $S(E)$ remain.

Using this approximation and the asymptotic expansion of the Whittaker
function,
we can obtain an analytical result for S(E),
which is not intended to be quantitative but will contain the
main energy dependence as $k \to 0$.  In the following pages we perform
the detailed derivation, but advise to follow onto Eq. (\ref{sfint3})
in case the reader is searching for a final result only.

\vspace{0.5cm}

First, we use the asymptotic expansion of the Whittaker function 
(at $r \to \infty$) \cite{gr80}:
\begin{equation}
W_{-\eta_{f},l_{f}+1/2} (2 \kappa\,r) \to
\frac{e^{-\kappa\,r}}{(2\,\kappa\,r)^{\eta_{f}}} \;\
\left\{ \sum\limits_{j=0}^{\infty}
\frac{(l_{f}-\eta_{f})_{j}\;
(l_{f}+\eta_{f}+j)_{j}}{j!\;(2\,\kappa\,r)^{j}}
\right\}\,
\label{asexpwh1}
\end{equation}
with $(l)_0=1$ and $(l)_j=1 \cdot l \cdot (l-1) \cdot ... (l-(j-1))$. 
Substituting in Eq. (\ref{sfctr2}) we arrive at:
\begin{eqnarray}
S(E)&=& {\cal A} \; \frac{e^{2\,\pi\,\eta_{i}}}{\mu_{ap}E}\,                          
\,(k\,r_{0})\,(k_{\gamma}\,r_{0})^{2\,L+1}\,
\nonumber\\
&&\times  \lbrack\, |\frac{1}{r_{0}^{L+1}}\,\sum\limits_{j=0}^{\infty}
\frac{(l_{f}-\eta_{f})_{j}\;
(l_{f}+\eta_{f}+j)_{j}}{j!}\, \int\limits_{r_{0}}^{\infty}\,
{\rm d}r\,r^{L}\,
e^{-\kappa\,r}\;   
\frac{1}{(2\,\kappa\,r)^{\eta_{f} + j}}
F_{l_{i}}(k,r)| \rbrack^{2}.
\label{sfctr3}
\end{eqnarray}
Before the integration over $r$ is performed,
by including the explicit behavior of the regular Coulomb function, we
can
estimate the analytical behavior of the integral in each term of the
asymptotic expansion,
\begin{equation}
T_{j}= \int\limits_{r_{0}}^{\infty}\,
{\rm d}r\,r^{L}\,
\frac{e^{-\kappa\,r}}{(2\,\kappa\,r)^{\eta_{f}+j}}
F_{l_{i}}(k,r). 
\label{asint1}
\end{equation}
The regular Coulomb function is given by:
\begin{equation}
F_{l_{i}}(k,r) =C_{l_{i}}(\eta_{i})\,(k\,
r)^{l_{i}+1}\,
e^{-i k r} \:\:_1F_1[l_{i}+1-i \eta_{i},2l_{i}+2;2\,i\,k\,r],
\label{eq:ss1}
\end{equation}
where $_1F_1[l_{i}+1-i \eta_{i},2l_{i}+2;2\,i\,k\,r]$ is a confluent
hypergeometric function (see Appendix,\, Eq. (\ref{kumerint1})),
and
\begin{equation}
C_{l_{i}}(\eta_{i})= 2^{l_{i}}\,e^{- \pi\,\eta_{i}/ 2}
\frac{|\Gamma(l_{i}+1+i \eta_{i})|}{\Gamma(2l_{i}+2)}.
\label{penetrfctr1}
\end{equation}
The asymptotic behavior of the Coulomb function as $r \to \infty$ is
well known:
\begin{equation}
F_{l_{i}}(k,r) \sim sin(k\,r - \frac{\pi}{2}\,l_{i}-
\eta_{i}\,\ln{2\,k\,r} + \sigma_{l_{i}}),
\label{coulas1}
\end{equation}
with $\sigma_{l_{i}}= arg \: \Gamma(l_{i} + 1 + i\eta_{i})$.
Although for the loosely bound
states we need to know the Coulomb wave function only for $r > r_{0}$,
the physical meaning of the $C_{l_{i}}(\eta_{i})$ coefficient
comes from the behavior of $F_{l_{i}}(k,r)$ at small $r$: 
\begin{equation}
F_{l_{i}}(k,r) \stackrel{r \to 0}{\approx}
C_{l_{i}}(\eta_{i})\,(k\,r)^{l_{i}}.    
\label{psir0}
\end{equation}
It is clear from Eq. (\ref{psir0}) that $C_{l_{i}}(\eta_{i})$ defines
the probability of finding a charged particle in the vicinity of $r=0$, 
under the influence of the Coulomb and centrifugal potentials.
Note that
\begin{eqnarray}
C_{l_{i}}^{2}(\eta_{i})&=& 2^{2l_{i}+1}\,\pi\,\eta_{i}\,
\frac{1}{e^{2\,\pi\,\eta_{i}}-1}\,\frac{1}{k^{2\,l_{i}}}\,
\frac{1}{[\Gamma(2\,l_{i}+2)]^{2}}
\prod_{j=1}^{l_{i}}(j^{2}\,k^{2}+ \eta_{f}^{2}\,\kappa^{2}).
\label{clet1}
\end{eqnarray}

In brief, we have replaced the lower integration limit $r_{0}$ by zero,
we have used the asymptotic expansion for the bound state
Eq. (\ref{asexpwh1})
and the explicit representation of the Coulomb scattering wave in
Eq. (\ref{eq:ss1}). After performing the integration,
the S-factor simplifies to:
\begin{eqnarray}
S(E)&=&
{\cal A} \; \frac{e^{2\,\pi\,\eta_{i}}}{\mu_{ap}E}\, \,
k^{2l_{i}+3}\, C_{l_{i}}^{2}(\eta_{i})\,k_{\gamma}^{2\,L+1}\,
\frac{1}{(2\,\kappa)^{2\,\eta_{f}}}\,                                  
\frac{1}{|(k - i\,\kappa)^{2-\eta_f + L +l_{i}}|^{2}}\,
\nonumber\\
&\times & 
| \sum_{j=0}^{\infty} \; \Upsilon_j \;i^{j}\,
\left(\frac{k - i\,\kappa}{2 \kappa}\right)^j \;
{}_2F_1[l_{i}+1-i\eta_{i},\, 2-\eta_f+ L +l_{i}-j,\, 2l_{i}+2,\,
\frac{2k}{k-i\,\kappa}]|^{2},
\label{sfint3}
\end{eqnarray}
 where the expansion coefficients are given by:
\begin{eqnarray}
\Upsilon_j  = (l_{f}-\eta_f)_{j} \; (l_{f} + \eta_f+j)_{j} \;
\frac{ \Gamma[2-\eta_f+ L + l_{i}-j]}{j!}.
\label{coef1}
\end{eqnarray}
We note that the expression for $S(E)$ in Eq. (\ref{sfint3})
is completely general.
In particular, the expression for the neutron overlap integral can
be obtained by calculating the limit $\eta_f,\,\eta_{i} \rightarrow 0$.

Substituting Eq. (\ref{clet1}) in Eq. (\ref{sfint3}) one arrives at:
\begin{eqnarray}
S(E)&=& {\cal A} \; 2^{2l_{i}+1}\,\pi \,\eta_{f}\,\kappa\,
\frac{1}{[\Gamma(2\,l_{i}+2)]^{2}} \;
\prod_{j=1}^{l_{i}}(j^{2}\,k^{2}+ \eta_{f}^{2}\,\kappa^{2}) \;\;\;\,
\nonumber\\     
& \times &\,                                  
k_{\gamma}^{2\,L+1}\,
\frac{1}{(2\,\kappa)^{2\,\eta_{f}}}\;
\frac{1}{|(k - i\,\kappa)^{2-\eta_f + L +l_{i}}|^{2}}\, 
| \sum_{j=0}^{\infty} \; \Upsilon_j \;i^{j}\,
\left(\frac{k - i\,\kappa}{2 \kappa}\right)^j \;
\nonumber\\     
&\times &\,       
{}_2F_1[l_{i}+1-i\eta_{i},\, 2-\eta_f+ L +l_{i}-j,\, 2l_{i}+2;\,
\frac{2k}{k-i\,\kappa}]|^{2}.
\label{sfint4}
\end{eqnarray}
This equation bears all the necessary features to investigate the
behavior of the S-factor approaching $E \to -\varepsilon$.

Beforehand, we make some obvious remarks concerning
the pole in S(E).  Taking into account the asymptotic behavior of 
$F_{l_{i}}(k,r)$ given in Eq. (\ref{coulas1}), we conclude that the 
singularity of $T_{j}$ on the physical
k-half-plane ($Im k > 0$) is due to the divergence of the
radial integral on the upper limit ($r \to \infty$) at $k \to 
i\,\kappa$. More explicitly, $T_{j}$ behaves as:
\begin{eqnarray}
T_{j} \propto \int\limits_{r_{0}}^{\infty}\,
{\rm d}r\,r^{L}\,
\frac{e^{-\kappa\,r}}{(r)^{\eta_{f}+j}}
e^{-i\,k\,r}r^{i\,\eta_{i}} \;.
\label{singint2}
\end{eqnarray}
When $k \rightarrow i \kappa$ then $i \eta_f \rightarrow \eta_f$.
Then, it is easy to verify that:
\begin{eqnarray}
T_{j} &\propto& (k - i\,\kappa)^{j-L-1} \;\mbox{ for } \; j < L+1 
\\
&\propto& \; \ln(k - i\,\kappa) \;\;\;\;\;\mbox{ for } \; j=L+1\;.
\label{sing1}
\end{eqnarray}
Note that $T_{j}$ is finite at $k= i\,\kappa$ for $ j > L + 1$.
Besides the square of the integral, S(E) contains also a factor
$k_{\gamma}^{2\,L+1}$ (see Eq.\ref{sfctr3}). Consequently it is the
term $j=0$
in the expansion that  generates the pole in $S(E)$, as first indicated
in
\cite{jenn98}. All higher order terms, $j > 0$, go to zero as $k \to
i\,\kappa\,$.
Let $A_j(E)$ be the agglomerate of all energy dependent factors in
Eq. (\ref{sfctr3}):
\begin{equation}
A_{j}(E) = \,\frac{1}{E}\,e^{2\,\pi\,\eta_{i}}\,(k\,r_{0})
(k_{\gamma}\,r_{0})^{2L+1} \;  |T_{j}| ^{2}\; .
\label{afctr1}
\end{equation}
Then its behavior as $k \to i \kappa$ (or $E \to -\varepsilon$) is:
\begin{eqnarray}
A_{j}(E) &\propto& \frac{1}{(E+ \varepsilon)^{1-j}}
\;\mbox{ for } \; j < L+1 \\
&\propto&
\lbrack \ln(E + \varepsilon) \rbrack^{2}\,(E + \varepsilon)^{2\,L+1}
\;\;\;\;\;\mbox{ for } \; j = L + 1\;.
\label{afctr2}
\end{eqnarray}
The term j=0 of Eq. (\ref{afctr1}) does not depend on $l_{f}$.
Hence, if this pole really did govern the behavior of the S-factor
at zero energy,  then $S(E)$ should increase rapidly as $E \to 0^+$,
independently of $l_{f}$.  We know from experiment that this is not the
case.  In fact, we will show that the rise of $S(E)$ at $E \to 0^+$
is dictated by higher order terms while the first order term,
the only one with a pole at $E=-\varepsilon$, actually decreases as $E
\to 0^+$.
To this end we calculate the behavior of $S(E)$  around threshold
for positive energies, $E \to 0^+$, but also for negative energies,
all the way down to $E \to -\varepsilon^+$.

This equation bears all the necessary features to investigate the
behavior
of the S-factor approaching $E \to -\varepsilon$. Let us begin with the
term j=0:
\begin{eqnarray}
S_{(0)}(E)&=& {\cal A} \; 2^{2l_{i}+1}\,\pi \, \eta_{f}\,\kappa\,
\frac{1}{[\Gamma(2\,l_{i}+2)]^{2}} \;
\prod_{j=1}^{l_{i}}(j^{2}\,k^{2}+ \eta_{f}^{2}\,\kappa^{2})\,
k_{\gamma}^{2\,L+1}\, \frac{1}{(2\,\kappa)^{2\,\eta_{f}}}                               
\nonumber\\    
&\times & \,                                  
\Upsilon_0^{2}\,|\frac{1}{(k - i\,\kappa)^{2-\eta_f + L +l_{i}}}\, 
 {}_2F_1[l_{i}+1-i\eta_{i},\, 2-\eta_f+ L +l_{i},\, 2l_{i}+2;\,
\frac{2k}{k-i\,\kappa}]|^{2}.
\label{sfj01}
\end{eqnarray}
In order to find the analytical properties of the astrophysical
factor around the pole,  $E \to -\varepsilon$, it is useful to perform
an
argument transformation relating the hypergeometric
functions with arguments $z$ and $1/z$, before taking the relevant
limit $k \to i \kappa$ \cite{gr80}:
\begin{eqnarray}
{}_2F_1[l_{i}+1-i\,\eta_{i},\, 2-\eta_f+ L +l_{i},\, 2l_{i}+2;\,
\frac{2k}{k-i\,\kappa}] \stackrel{k \to i\,\kappa}{\sim} \nonumber \\
c_{1}\,(k - i\,\kappa)^{l_{i} + 1 - i\,\eta_{i}} +
c_{2}\,(k - i\,\kappa)^{2 - \eta_{f} + L + l_{i}}.
\label{hypgs1}
\end{eqnarray}
Using this result in Eq. (\ref{sfj01}) we conclude that the leading
singular term for $k \to i\,\kappa$ is:
\begin{equation}
S_{(0)}(E) \stackrel{E \to -\varepsilon}{\propto} \frac{1}{(E +
\varepsilon)},
\label{sfcsngbh1}
\end{equation}
whichever $l_{i},\,l_{f}$ or $L$, consistent with the conclusions drawn
from Eq. (\ref{afctr1}).

It is important to realize that the behavior of the hypergeometric
function ${}_2F_1[l_{i}+1-i\eta_{i},\, 2-\eta_f+ L +l_{i},\,
2l_{i}+2;\,
z=2k/(k-i\,\kappa)]$ is very different around the pole ($z \to \infty$)
compared to its behavior around threshold ($ z \to 0$).
The analytical expression of $S(E)$  at threshold will
been derived in Section \ref{thresh}. 

${}_2F_1[l_{i}+1-i\eta_{i},\,2-\eta_f+ L +l_{i},\, 2l_{i}+2;\, 
2k/(k-i\,\kappa)]$ 
is an analytic function in the finite $k$-plane 
(except for $k = \pm i\,\kappa$).
Hence, assuming that the cuts drawn from branching 
points $k = \pm i\,\kappa$ go to infinity,  
we can make an analytic continuation of 
${}_2F_1[l_{i}+1-i\eta_{i},\,2-\eta_f+ L +l_{i},\, 2l_{i}+2;\, 
2k/(k-i\,\kappa)]$ into the complex $k$-plane. 
Using Eq. (\ref{sfj01}) we arrive at an expression for the S-factor 
in the vicinity of the pole. 
Note that the continuation of S(E) will not necessarily be analytical at 
$k=0$, as the modulus operation may introduce a discontinuity.
For $k=i\,p$ and $E=-{p^{2}}/2\,\mu_{ap}$, Eq. (\ref{sfj01}) takes the
form
\begin{eqnarray}
S_{(0)}(-|E|)&=& {\cal A} \; 2^{2l_{i}+1}\,\pi \,\eta_{f}\,\kappa\,
\frac{{\Upsilon_0}^2}{[\Gamma(2\,l_{i}+2)]^{2}} \;
\prod_{j=1}^{l_{i}}(j^{2}\,k^{2}+ \eta_{f}^{2}\,\kappa^{2})\,
\frac{k_{\gamma}^{2\,L+1}}{(2\,\kappa)^{2\,\eta_{f}}}\,
\nonumber\\
&&\times\, \frac{1}{( \kappa-p)^{4-2\,\eta_f + 2\,L +2\,l_{i}}}\,
({}_2F_1[l_{i}+1-\eta_{i}^{'},\, 2-\eta_f+ L +l_{i},\,
2l_{i}+2,\, \frac{2p}{p - \kappa}])^{2},
\label{sfj02}
\end{eqnarray}
where $\eta_{i}^{'} = Z_{a}\,e^{2}\,\mu_{ap}/p$.

\begin{figure}[h!]
\centerline{
        \parbox[t]{0.5\textwidth}{
\centerline{\psfig{figure=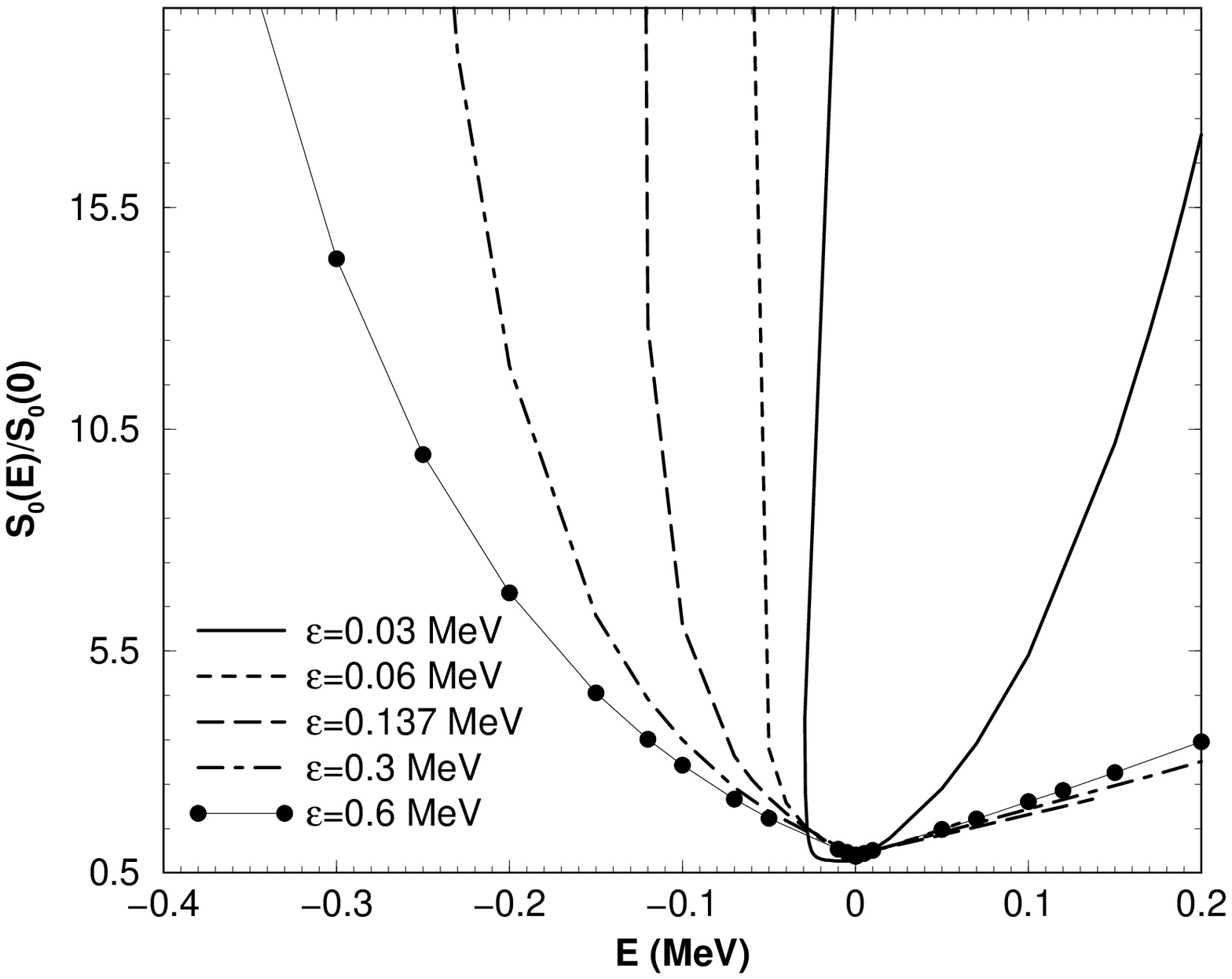,width=0.5\textwidth}}
        \caption{$S_{0}(E)$ for the $l_{i}=0,\, \to l_{f}=1$
        proton capture, for a set of binding energies of the final
        state. Masses and charges are the same as for 
$^7$Be(p,$\gamma$)$^8$B.
        $S_{0}(E)$ is normalized to unity at zero energy.}
\label{fig_S0}}
}
\end{figure}

In Fig.\ref{fig_S0} we present the energy
behavior of $S_{(0)}(E)$  for $a(p,\gamma)b$, for masses $A_a=7$ and
$A_b=8$\,
and angular momentum $l_{i}=0,\,l_{f}=1$  and $L=1$.
It mimics ${}^{7}{\rm Be}(p, \gamma){}^{8}{\rm B}$, but a different set
of
proton binding energies are used. The energy interval is
$-\varepsilon < E \leq \varepsilon$. All $S_{(0)}(E)$ are normalized to 
unity
at
zero energy.
It is obvious that, even for the smallest binding energy, in which the
pole is closest to threshold, $S_{(0)}(E)$ decreases as $E \to
0^+$. Whenever there is a  rise of S(E) around zero, it can only
be due to the sum effect of higher order terms in Eq. (\ref{sfint4}),
and not the pole.


\section{Threshold Behavior of the S-factor}
\label{thresh}

In this section, we look at the dependence of the derivative
of $S(E)$ on $l_i$ and $l_f$, and analyze the competition
between the remnant initial Coulomb barrier and the trace of
the singularity $E=-\varepsilon$, discussed in the previous section.
We also investigate the dependence of $S(E)$ on charge
and binding energy.

Using the integral representation of the initial scattering state,
and the final bound state, in the asymptotic approximation, one
can deduce an integral expression for the S-factor close to threshold.
This derivation is presented in detail in Appendix \ref{app1} and
the result we obtain is:
\begin{eqnarray}
S(E)&=& {\cal B} \, k_{\gamma}^{2\,L+1}\,
\frac{\kappa^{2l_{f}+1}}{\eta_{f}} \,
\prod_{j=0}^{l_{i}}(j^{2}\,k^{2}+ \eta_{f}^{2}\,\kappa^{2}) \;
\nonumber\\
&&\times \bigg| \int\limits_{0}^{\infty}\,{\rm d}s\,
{\cal S}(s) \; \lim_{\epsilon \to 0}\,
\frac{\rm d^{m}}{\rm d\, \epsilon^{m}}\,
\frac{1}{\lbrack (\kappa\,(1+2\,s)+
\epsilon)^{2}+k^{2}\rbrack^{l_{i}+1}}\,
e^{2\,\eta_{i}\arctan \frac{k}{\kappa(1+2\,s)+ \epsilon}}\bigg|^{2},
\label{sfctrf2}
\end{eqnarray}
where an energy independent constant ${\cal B}$ has been introduced
(Eq.\ref{constB}),
${\cal S}(s)$ is the spectral function defined in Eq. (\ref{intSs})
and $m=l_{f}+L+1-l_{i}$.

We mentioned that, by taking only the first term of the
expansion of the Whittaker function defined in Eq. (\ref{asexpwh1})
we can arrive at a similar expression for $S_{(0)}$ where ${\cal S}(s)$
should be replaced by ${\cal S}_{0}(s)$ defined in Eq. (\ref{intSs0}).
Although $S_{(0)}(E)$ has a pole at
$k=i\,\kappa$, as $k \to +0$ its first
derivative $S'_{(0)}(E) >0$ for all binding energies, as can be 
verified
in Fig.\ref{fig_S0}. 
The only difference between  $S(E)$ and $S_{(0)}(E)$
is in the spectral functions for
$W_{-\eta_{f},l_{f}+1/2}(2\,\kappa\,r)$, Eq. (\ref{intwhit1}),
and $W_{-\eta_{f},l_{f}+1/2}^{(0)}(2\,\kappa\,r)$,
Eq. (\ref{intwhitf0}). The spectral functions   
at $s \to \infty$ are, from Eq. (\ref{intSs}) and Eq. (\ref{intSs0}),
\begin{equation}
{\cal S}(s)=[s(1+s)]^{l_{f}}\,\big[\frac{s}{1+s}\big]^{\eta_{f}}
\stackrel
{s \to \infty}{\approx} s^{2\,l_{f}} \;\mbox{ and } \;\hfill  \;\;
{\cal S}_{(0)}(s)=s^{l_{f}+\eta_{f}}.
\label{spctrf1}
\end{equation}
There are only two competing factors here: the centrifugal barrier
($l_f$)
and the Coulomb barrier ($\eta_f$) of the final-state.
Three situations may occur:\\
i) for $l_{f}=\eta_{f}$, the S-factor is represented exactly by the
first term: $S(E) \equiv S_{(0)}(E)$;\\
ii) for $l_{f} < \eta_{f}$,
${\cal S}_{(0)}(s)/{\cal S}(s) \stackrel{s \to \infty}{\longrightarrow}
\infty$:
the contribution of the large $s$ values is more
pronounced for $S_{(0)}(E)$ than for $S(E)$;\\
iii) for $l_{f} > \eta_{f}$,
${\cal S}(s)/{\cal S}_{(0)}(s) \stackrel{s \to \infty}{\longrightarrow}
\infty$:
the contribution of the large $s$ values is more
pronouncened for $S(E)$.  \\
In brief, if the final-state Coulomb barrier
prevails, ${S_{(0)}}'(0) > S'(0)$ otherwise,
if the final-state centrifugal barrier prevails, $S'(0) > S_{(0)}'(0)$.

Only two cases need to be considered for dipole transitions $L=1$:
$m=1$ if $l_{i}=l_{f}+1$
and $m=3$ if $l_{i}=l_{f}-1$.
We consider them separately.

\subsection{{\mathversion{bold} $m=1$} when 
{\mathversion{bold}$l_{i}=l_{f}+1$}.}

After taking the required derivative in Eq. (\ref{sfctrf2}),
the expression for $S(E)$  is
\begin{eqnarray}
S(E)&=& 4 {\cal B} \; \eta_{f}\,                     
\frac{k_{\gamma}^{2\,L+1}}{\kappa}\,          
\prod_{j=1}^{l_{i}}(j^{2}\,k^{2}+ \eta_{f}^{2}\,\kappa^{2})\,I^2(k^2),         
\label{sfctrf3}
\end{eqnarray}
with
\begin{equation}
I(k^2) = \int\limits_{0}^{\infty}\,{\rm d}s\,
\underbrace{\lbrack s(1+s) \rbrack^{l_{f}} \, \lbrack \frac{s}{1+s}
\rbrack^{\eta_{f}}}_{{\cal S}(s)}\, 
\underbrace{\frac{(l_{i}+1)\,(1+2\,s)+ \eta_{f}}{[(1+2\,s)^{2}
+  E/\varepsilon]^{l_{i}+2}}}_{f_1(s,E)} \;
\underbrace{e^{2\,\eta_{i}\arctan \frac{k}{\kappa(1+2\,s)}}}_{f_2(s,k)}
\;.
\label{intk2}        
\end{equation}

\vspace{0.5cm}

Let us first make some general qualitative remarks concerning the
behavior of the integrand near threshold $E\rightarrow 0^+$.
These will be useful later.
In all cases considered, the integrand peaks at $s_m >0$.
This peak occurs for a larger $s_m$, for $l+1 \to l$ than for $l \to
l-1$,
reflecting the effect of the centrifugal barriers in the initial and
final state. In addition, the peak of the integrand also shifts to
larger s-values
when the energy increases.

Remarkably, the energy dependent part of the
S-factor depends on $E$ rather than on $k$ and can be written as
a positive definite function:
\begin{equation}
F(E)= \bigg[k_{\gamma}^{3}\,
\prod_{j=1}^{l_{i}}(j^{2}\,k^{2}+
\eta_{f}^{2}\,\kappa^{2})\bigg]\,I^{2}(k^{2})\;.
\label{fE1}
\end{equation}
The part in square brackets decreases monotonically at $ E \to 0$
(positive derivative) while the integral $I(k^{2})$ increases as $E \to
0$
(negative derivative).
The sign of the first derivative of $S(E\rightarrow 0^+)$,
or in other words, the low-energy behavior of the S-factor, is defined
by the sign of the derivative of $F(E)$.

Next, we focus our attention on $I(k^{2})$. 
The integrand of $I(k^{2})$ is a positive definite
function of $s$ and $k$ and the two functions in the integrand
of Eq. (\ref{intk2}), $f_{1}(k^{2},\,s)$ and $f_{2}(s,\,k)$,
decrease monotonically with increasing energy. Now the interplay
of the various ingredients becomes more transparent.
Clearly, $f_{2}(s,\,k^{2})$ is a remnant of the initial
Coulomb barrier, as well as the second term of function
$f_{1}(s,\,k^{2})$,
$\eta_{f}/\lbrack (1+2\,s)^{2} +  E/\varepsilon\rbrack^{l_{i}+2}$.

Note that, as $l_{i}=l_{f}+1$, the initial orbital cannot be an $s$-wave.
As  increasing $l_{i}, l_{f}$ shifts the peak of the
integrand to higher values of $s$, the integrand, and,
hence, $I(k^{2})$, becomes less sensitive to energy variations.
If $I'(E)< 0$ increases as $l_{i}=l_{f}+1$ increases, the factor
$k_{\gamma}^{2\,L+1}\,\prod_{j=1}^{l_{i}}(j^{2}\,
k^{2}+ \eta_{f}^{2}\,\kappa^{2})\,$ eventually prevails
leading to the positive derivative $S'(E)$.

In addition, increasing $\varepsilon$ shifts the pole away
from threshold, such that $I(k^{2})$ becomes
less varying with energy around threshold. So, increasing the binding
energy, decreases $|I'(k^2=0)|$.

\vspace{0.5cm}

We can derive an expression for the S-factor
near threshold factorizing the energy dependence.
First we rewrite $I(k^{2})$ as
\begin{eqnarray}
I(k^{2})&=& \,\frac{1}{(E  + \varepsilon)^{2}}\,{\tilde I}(k^{2}),
\label{ikpl1}   \\
&&\stackrel{E \to 0^{+}}{=} I(0) + I'(0)\,E +
O([E/\varepsilon]^{2}),
\label{ik-exp}
\end{eqnarray}
where ${\tilde I}(0) \not= 0$.

Let us estimate the impact of the pole singularity on $I(k^{2})$
in the physical region at $E \to 0^{+}$.
In the case of a pure second order
pole, ${\tilde I}(k^{2}) \equiv const,\;\;-\varepsilon \; I'(0)/I(0)= 
2$
(the order of the pole). Otherwise, the Taylor expansion holds:
${\tilde I}(k^{2})= {\tilde I}(0)+
{\tilde I}'(0)\,E + O((E/\varepsilon)^{2})$, where
${\tilde I}'(0)/{\tilde I}(0) > 0$. Then we get
$-\varepsilon \; I'(0)/I(0)= [2-\varepsilon\,{\tilde I}'(0)/{\tilde 
I}(0)]$.
If $[(\varepsilon\,{\tilde I}'(0)/{\tilde I}(0)] > 0$, which 
corresponds
to our case, the power of the singularity in $I(k^{2})$ as $E \to 
0^{+}$ 
is weaker than $2$, {\it i.e.} $\varepsilon\,I'(0)/I(0) > - 2$.

The explicitly terms appearing in Eq. (\ref{ik-exp}) can be derived:
\begin{eqnarray}
I(0) & = &\int\limits_{0}^{\infty}\,{\rm d}s\,
\lbrack s(1+s) \rbrack^{l_{f}} \, \lbrack \frac{s}{1+s}
\rbrack^{\eta_{f}}\,\frac{(l_{i}+1)\,(1+2\,s)+ \eta_{f}}
{(1+2\,s)^{2\,l_{i}+4}}\,
e^{2\,\eta_{f}/(1+2\,s)} \;\mbox{  and}
\nonumber \\
I'(0) & = & -\frac{1}{\varepsilon}\,\int\limits_{0}^{\infty}\,{\rm
d}s\,
\lbrack s(1+s) \rbrack^{l_{f}} \, \lbrack \frac{s}{1+s}
\rbrack^{\eta_{f}}\,\frac{(l_{i}+1)\,(1+2\,s)+ \eta_{f}}
{3\,(1+2\,s)^{2\,l_{i}+6}}             \nonumber\\
&  & \mbox{\hspace{4cm}}
[3\,(l_{i}+2)\,(1+2\,s) + 2\,\eta_{f}]\,e^{2\,\eta_{f}/(1+2\,s)}.
\label{iz1}
\end{eqnarray}

Finally,  we can also write down the
threshold behavior of $S(E)$, as
$S(E) \stackrel{E \to 0}{=} S(0) + S'(0)\,E + O([E/\varepsilon]^{2})$
where:
\begin{equation}
S(0)= 4 {\cal B}\;\varepsilon^{3}\, \kappa^{2l_{f}+3}\,
\eta_{f}^{2l_i+1}\,\kappa^{2 l_i}\,I^{2}(0), \label{sz1}
\end{equation}
and
\begin{equation} 
S'(0)= 4 {\cal B}\, \varepsilon^{2}\, \kappa^{2l_{f}+3}\, 
\eta_{f}^{2 l_i+1}\,\kappa^{2 l_i}\,I^{2}(0)
\bigg[ 3 + \frac{1+4+..+l_{i}^{2}}{\eta_{f}^{2}}
+ 2\,\frac{\varepsilon\,I'(0)}{I(0)} \bigg].
\label{sderz1}
\end{equation}
From Eq. (\ref{sderz1}), it is clear that the sign of the derivative
of the S-factor at $E \to 0$ is
defined by the sign of
$\bigg[ 3 + (1+4+..+l_{i}^{2})/\eta_{f}^{2} -
2\,\varepsilon\,|I'(0)/I(0)| \bigg]$.
Note that the larger is $l_{i}$, 
the larger becomes the positive term in this factor. 

We now illustrate these ideas with a few examples which relate to
physical
cases.

\medskip

\noindent
{\bf (i)} {\mathversion{bold}$l_{i}=1,\,l_{f}=0$.}

\begin{figure}[htb]
\centerline{
        \parbox[t]{0.5\textwidth}{
\centerline{\psfig{figure=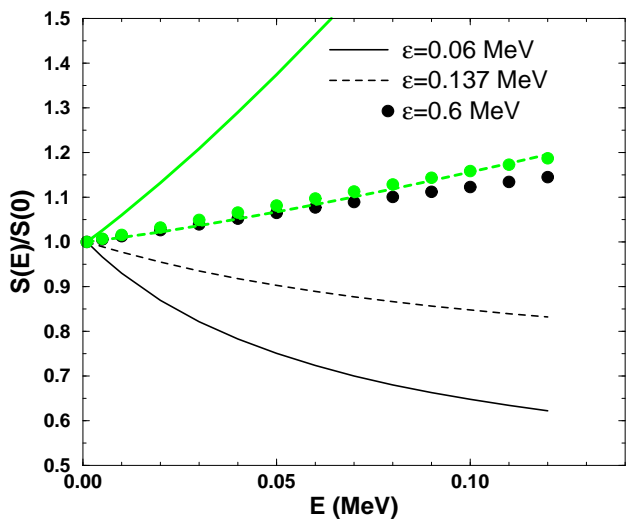,width=0.5\textwidth}}
        \caption{Proton capture $l_{i}=1,\to l_{f}=0$
        for a set of binding energies: $S(E)$ (dark lines) versus
        $S_{(0)}(E)$ (grey lines). Masses and charges are the same as
        for $^7$Be(p,$\gamma$)$^8$B. All
        S-factors are normalized to unity at zero energy.}
\label{fig_li1lf0}}
}
\end{figure}

In Fig.\ref{fig_li1lf0}
we present the behavior of $S(E)$ for the $l_{i}=1 \to l_{f}=0$ capture
${}^{7}{\rm Be}(p, \gamma)"{}^{8}{\rm B}"$, spanning a few values
for the binding energy of $"{}^{8}{\rm B}"$.
All $S(E)$ are normalized to unity at $E=0$.
The smaller the binding energy, the larger the increase of $S(E \to
0)$.
The centrifugal barrier in the initial state is not strong
enough to overcome the negative derivative
caused by the remnant Coulomb ($\eta_i$) and the singularity at
$E=- \varepsilon$. In addition there is the dependence on the
spectral function $\lbrack \frac{s}{1+s} \rbrack^{\eta_{f}}$,
where $\eta_f$ depends of the inverse square root of the
binding energy.
For $\varepsilon=0.06, \; 0.137,\;  0.35, \; 0.6$ MeV the Coulomb
parameters
are $\eta_f=2.4, \; 1.6, \; 1.0, \; 0.76$, respectively.
Replacing $l_{f}=0$ in Eq. (\ref{iz1}) we see that,
for larger $\eta_f$, a significant contribution
to the integral comes from smaller s-values,
reducing  $\varepsilon\,I'(0)/I(0)<0$ and $1/{\eta_{f}}^{2}$, so
that $\bigg[ 3 + 1/\eta_{f}^{2}+ 2\,\varepsilon\,I'(0)/I(0) \bigg]$
becomes negative. 
On the other hand, for smaller $\eta_f$, which corresponds to the pole
at $E=-\varepsilon$ moving further away from threshold,
the slope $\varepsilon\,I'(0)/I(0)$ and $1/{\eta_{f}}^{2}$ increase 
and,
correspondingly,
$\bigg[3 + 1/\eta_{f}^{2}+ 2\,\varepsilon\,I'(0)/I(0) \bigg]$
can become positive.

We also compare $S(E)$ with $S_{(0)}(E)$, obtained
retaining only the first term of the Whittaker expansion
(Fig.\ref{fig_li1lf0}).
Contrary to $S'(E)$ we find that all calculated cases have
positive $S'_{(0)}(0^+)$. This confirms that the negative
derivative of $S(E \to 0)$ is not due to the nearby pole as concluded
in Section \ref{sect_1}.
As $\eta_{f} >l_{f}=0$, ${S_{(0)}}'(0)$ is always larger than $S'(0)$. 

\medskip

\noindent
{\bf (ii)} {\mathversion{bold}$ l_{i}=2,\,l_{f}=1.$}

\begin{figure}[htb]
\centerline{
        \parbox[t]{0.5\textwidth}{
\centerline{\psfig{figure=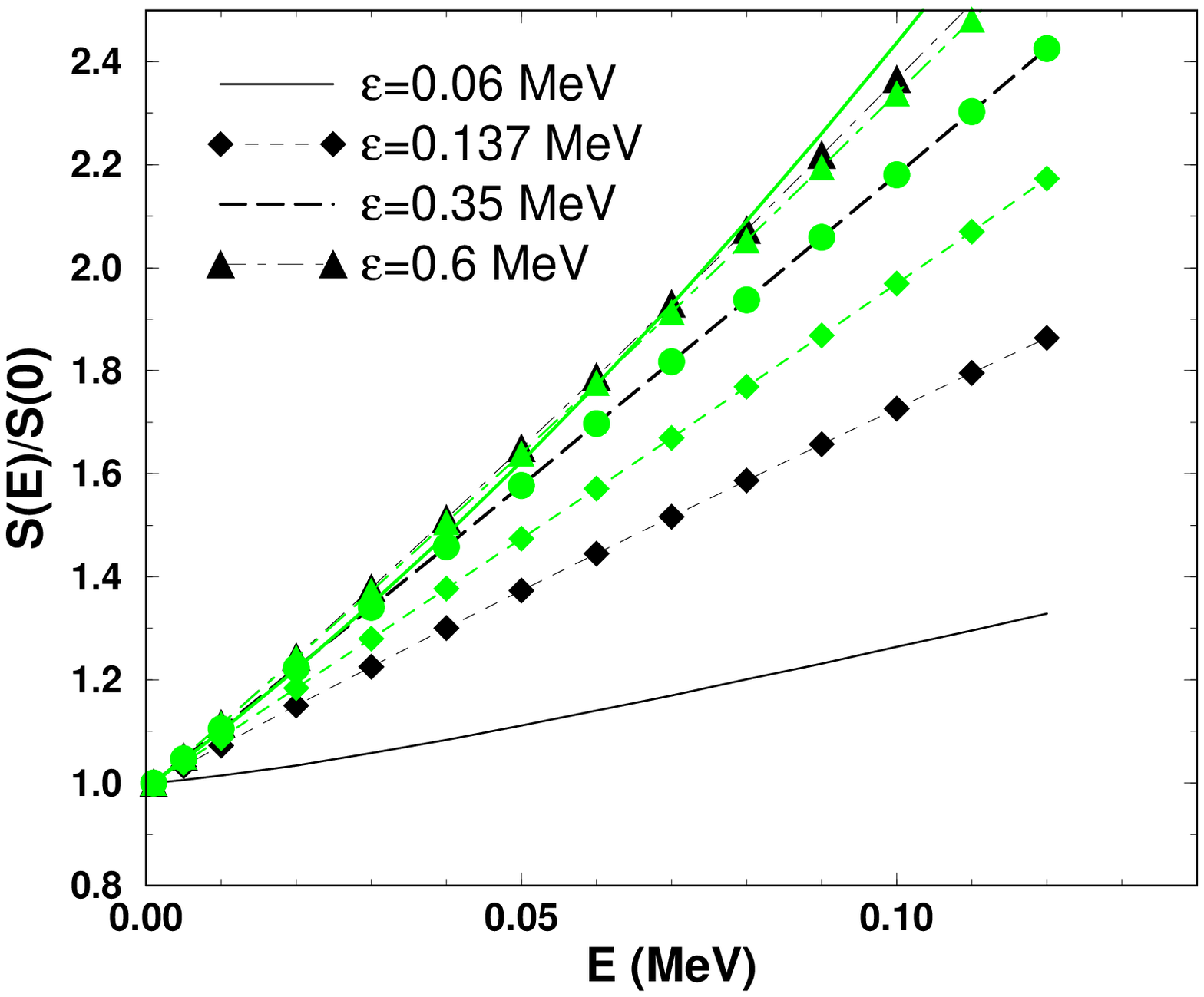,width=0.5\textwidth}}
        \caption{Proton capture $l_{i}=2,\to l_{f}=1$
        for a set of binding energies: $S(E)$ (dark lines) versus
        $S_{(0)}(E)$ (grey lines). For $\varepsilon = 0.09$ MeV, 
$S_{(0)}(E)$
        is plotted in circles to distinguish from $S(E)$.
        Masses and charges are the same as for $^7$Be(p,$\gamma$)$^8$B. 
All
        S-factors are normalized to unity at zero energy.}
\label{fig_li2lf1}}
}
\end{figure}

In Fig. \ref{fig_li2lf1} are the S-factors for the
$l_{i}=2 \to l_{f}=1$ capture
${}^{7}{\rm Be}(p, \gamma)"{}^{8}{\rm B}"$,
assuming a set of binding energies for $"{}^{8}{\rm B}"$.
Strikingly, all binding energies produce a positive derivative
for $S(0)$.
In this case the centrifugal barriers in the initial and final states
overcome the remnant Coulomb barrier and the pole at $E=-\varepsilon$.
Thus, even at the lowest binding energy, $\varepsilon=0.06$ MeV,
$S'(0)$ is positive. When the binding energy
increases, the pole moves away from threshold
(decreasing $\varepsilon\,I'(0)/I(0)$) and $\eta_f$ decreases,
which means that
$\bigg[ 3 + 5/\eta_{f}^{2}+ 2\,\varepsilon\,I'(0)/I(0) \bigg]$
becomes larger, as well as $S'(0)$.
The general behavior for $l_{i} > 2 \to l_f> 1$ is identical. 

Fig. \ref{fig_li2lf1} also shows the behavior of $S_{(0)}(E)$,
represented by the grey lines. This figure is a clear illustration of
the points mentioned in the beginning of this section. 
For the cases $\varepsilon=0.06$ and
$\varepsilon=0.137$ MeV, $\eta_{f} > l_{f}(=1)$
and then ${S_{(0)}}'(0) > S'(0)$.
The binding $\varepsilon=0.35$ MeV corresponds to $\eta_{f}=1$
for which  ${S_{(0)}}(0) \equiv S(0)$.
Finally, for $\varepsilon=0.6$ MeV, 
$\eta_{f} < l_f$ and consequently ${S_{(0)}}'(0) < S'(0)$.


\subsection{{\mathversion{bold} $m=3$} when
{\mathversion{bold}$l_{i}=l_{f}-1$}.}

Directly from Eq. (\ref{sfctrf2}) we get
\begin{eqnarray}
S(E)& = & {\cal B} \; k_{\gamma}^{3}\,
\frac{\kappa^{2l_{f}+1}\,}{\eta_{f}^{2}}\,
\prod_{j=0}^{l_{i}}(j^{2}\,k^{2}+ \eta_{f}\,\kappa^{2})    
\nonumber\\
&\times & \bigg|\lim_{\epsilon \to 0}\,
\int\limits_{0}^{\infty}\,{\rm d}s\,
{\cal S}(s)\, \frac{\rm d^{3}}
{\rm d\, \epsilon^{3}}\,
\frac{1}{\lbrack (\kappa\,[1+2\,s] + \epsilon)^{2} + 
k^{2}\rbrack^{l_{i}+1}}\,
e^{2\,\eta_{i}\arctan \frac{k}{\kappa(1+2\,s)+ \epsilon}}
\bigg|^{2}.
\label{sfctrf4}
\end{eqnarray}
The explicit expression for the third derivative of the integrand
is  tedious, but the analysis is in all similar to the
case with $m=1$. The energy dependence of the S-factor
has the same form as Eq. (\ref{fE1}),
\begin{equation}
F(E)= \bigg[k_{\gamma}^{3}\,\prod_{j=0
}^{l_{i}}(j^{2}\,k^{2}+ \eta_{f}^{2}\,\kappa^{2})\bigg]\,
I_3^{2}(k^{2}),      
\label{fE2}
\end{equation}
only the integral changes due to the third derivative, which we here
represent as $I_3(k^2)$.
We keep in mind that the initial state now can be an $s$-wave
($l_{i}=0$).
$S(0)$ and $S'(0)$ are given by expressions identical to
Eq. (\ref{sderz1}),
replacing $I(0)$ and $I'(0)$ by $I_3(0)$ and $I'_3(0)$, which have
rather
complicated explicit forms. We apply these expressions for the
two most common cases found in Astrophysics: transitions $0 \to 1$ and
$1 \to 2$.

\medskip

\noindent
{\bf (i){\mathversion{bold} $l_{i}=0 \to l_{f}=1$.}}

\begin{figure}[htb]
\centerline{
        \parbox[t]{0.5\textwidth}{
\centerline{\psfig{figure=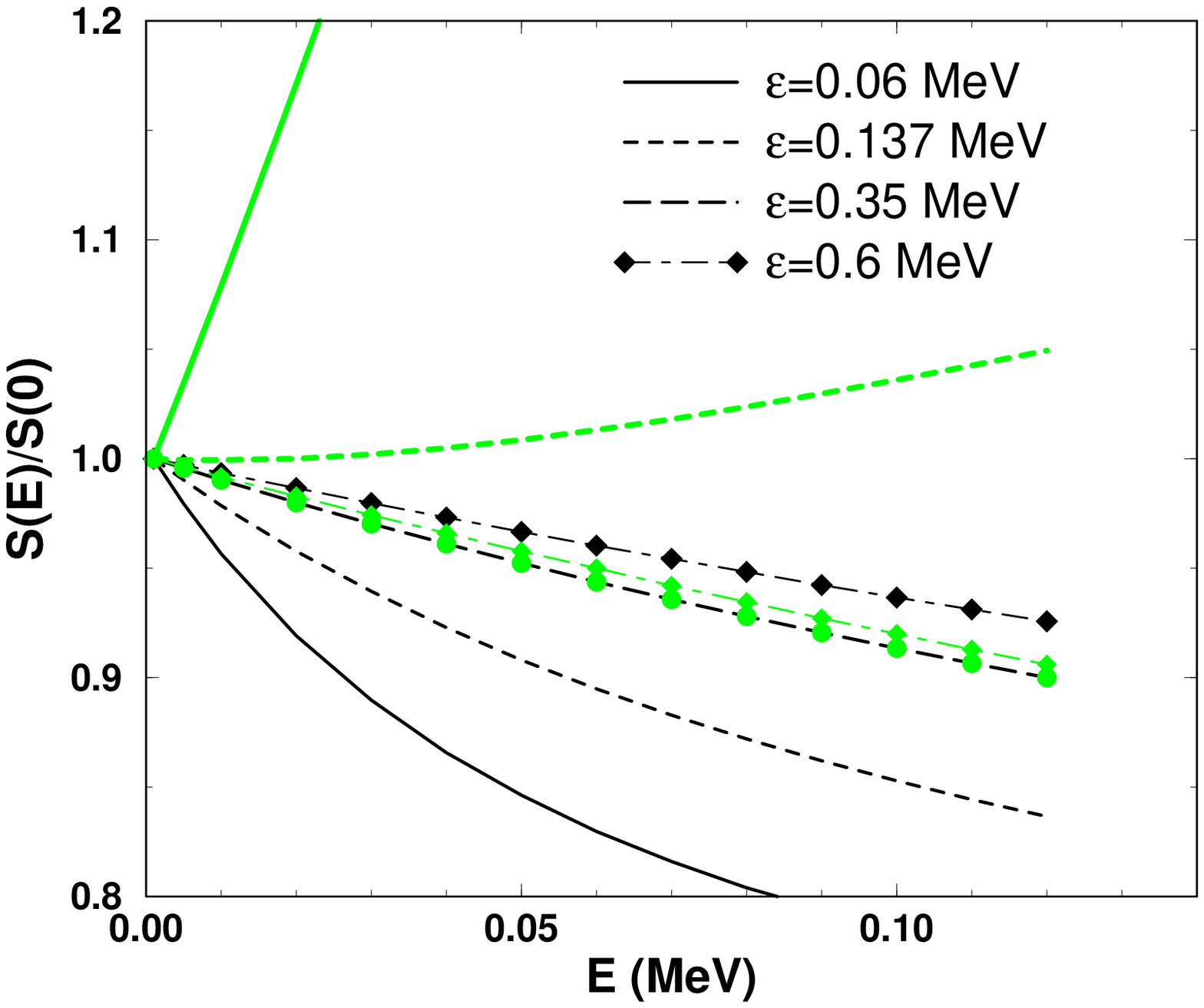,width=0.5\textwidth}}
        \caption{Proton capture $l_{i}=0,\to l_{f}=1$
        for a set of binding energies: $S(E)$ (dark lines) versus
        $S_{(0)}(E)$ (grey lines). For $\varepsilon = 0.35$ MeV, 
$S_{(0)}(E)$
        is plotted in circles to distinguish from $S(E)$.
        Masses and charges are the same as for $^7$Be(p,$\gamma$)$^8$B. 
All
        S-factors are normalized to unity at zero energy.}
\label{fig_li0lf1}}
}
\end{figure}

Similarly to what was done in the previous subsection,
we consider $S(E)$ for the $l_i=0 \to l_f=1$ capture
in the case of ${}^{7}{\rm Be}(p, \gamma)"{}^{8}{\rm B}"$
taking a set of binding energies for $"{}^{8}{\rm B}"$. 
Results are presented in Fig. \ref{fig_li0lf1}.
In this case, there is no initial centrifugal 
barrier and the centrifugal barrier in the final state alone cannot
overcome the rising of $I_{3}^{2}(E \sim 0)$ due to the pole
and the remnant Coulomb.  From Eq. (\ref{intk2}) one can verify that
the asymptotics of spectral function which goes as $s^{2}$ (see
Eq.\ref{spctrf1}),
trends to push the peak  of the integrand to high $s$-values.
This would make $I_3(k^2)$ less sensitive to variation of $E$,
yet it is the third derivative that enhances this sensitivity,
creating the net effect.

By comparing Fig. \ref{fig_li1lf0} and Fig. \ref{fig_li0lf1},
one realizes that $S'(0)$ increases with binding energy for both
cases. While for $l_i=1 \to l_f=0$ the sign of $S'(0)$ becomes
positive, all  $l_i=0 \to l_f=1$ show  $S'(0)<0$, suggesting
that the centrifugal barrier has a stronger impact in
the initial state.

In Fig. \ref{fig_li0lf1} the comparison between $S(E)$ and
$S_{(0)}(E)$ (grey lines) is also done. Whenever
$\eta_{f} > l_{f}$ ($\varepsilon= 0.06, \; 0.137$ MeV)
${S_{(0)}}'(0) > S'(0)$; for $\eta_{f}=l_{f}$ ($\varepsilon= 0.35$ MeV)
the two functions coincide and for $\eta_{f} < l_{f}$ ($\varepsilon=
0.6$ MeV)
${S_{(0)}}'(0)< S'(0) $.

\medskip

\noindent
{\bf (ii) {\mathversion{bold} $l_{i}=1 \to l_{f}=2$.}}

\begin{figure}[htb]
\centerline{
        \parbox[t]{0.5\textwidth}{
\centerline{\psfig{figure=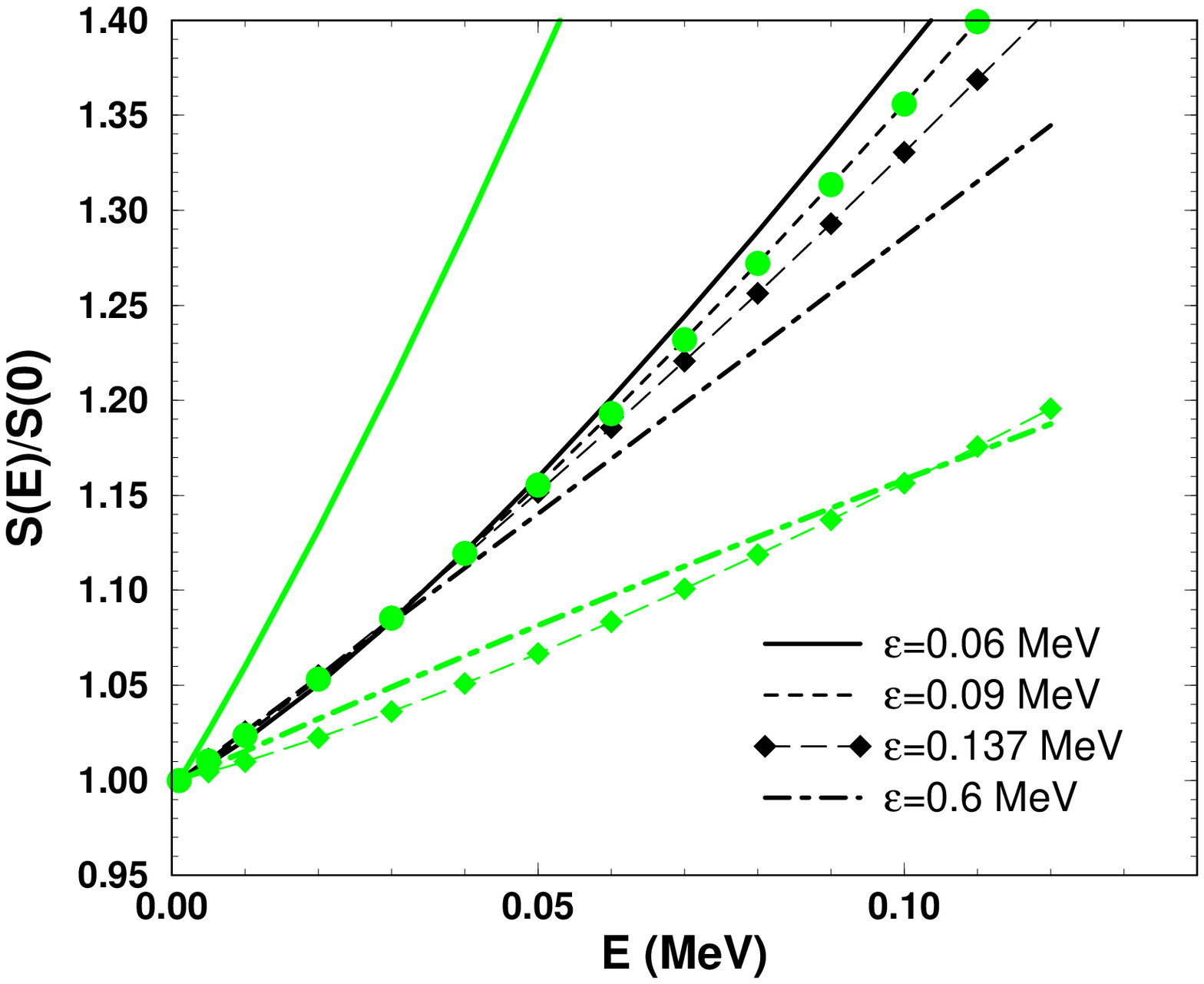,width=0.5\textwidth}}
        \caption{Proton capture $l_{i}=1,\to l_{f}=2$
        for a set of binding energies: $S(E)$ (dark lines) versus
        $S_{(0)}(E)$ (grey lines). For $\varepsilon = 0.09$ MeV, $S_{(0)}(E)$
        is plotted in circles to distinguish from $S(E)$.
        Masses and charges are the same as for $^7$Be(p,$\gamma$)$^8$B. All
        S-factors are normalized to unity at zero energy.}
\label{fig_li1lf2}}
}
\end{figure}

In Fig. \ref{fig_li1lf2} are the S-factors for
the $l_{i}=1 \to l_{f}=2$ capture for the reaction
${}^{7}{\rm Be}(p, \gamma)"{}^{8}{\rm B}"$ where
$"{}^{8}{\rm B}"$ is bound by a similar set of binding energies
(we chose instead of $\varepsilon=0.35$ MeV, $\varepsilon=0.09$ MeV for
which $\eta_f=2$ matching $l_f$).
As in Fig. \ref{fig_li2lf1}, $S'(0)>0$ for all cases.
Clearly here the centrifugal barriers are enough to overcome
the influence of the remnant Coulomb barrier and the pole at
$E=-\varepsilon$.
Note that the factor preceding the integral is stronger
in this case: $F(E, 1 \to 2)=k_{\gamma}^3 ( k^2 + \eta_f^2 \kappa^2)
I^2_3(k^2)$,
whereas  $F(E, 0 \to 1)=k_{\gamma}^3 I^2_3(k^2)$.
This factor is responsible for the decrease of S(E) as $E \to 0$.

A subtle feature appears that makes this example contrast with
all previously considered cases: when the binding energy increases
(the pole moves away from the threshold), $S'(E)$ decreases.
To understand this property, one would have to develop the
third derivative explicitly.
The safe conclusion that can surely be drawn is
that the strong centrifugal barrier washes out the dependence on 
binding
energy.

The grey lines in Fig.\ref{fig_li1lf2} show the $S_{(0)}(E)$
behavior. Identical conclusions to those obtained in (i)
can be drawn from the results:
when $\eta_f=l_f$ $S(E)$ coincides with $S_{(0)}(E)$ (grey circles);
when $l_{f} < \eta_{f}$ then ${S_{(0)}}'(0) > S'(0)$ (solid lines),
the opposite happening when $l_{f} > \eta_{f}$ (dashed and dot-dashed
lines).


\subsection{Charge dependence}

The energy dependence of the S-factors is strongly influenced by the
charge of the capturing nucleus. It is through $\eta_i$, the remnant
Coulomb
in the initial state, and $\eta_f$, the Coulomb parameter of the final
state, that this dependence comes in. In order to illustrate the
relative interplay between these parameters we chose again the typical
case
$"{}^{7}{\rm Be}"(p, \gamma)"{}^{8}{\rm B}",\;\varepsilon =0.137$ MeV,
where both $l_{i}=0 \to l_{f}=1$ and $l_{i}=1 \to l_{f}=2$
transitions are considered
and the charge of the capturing nucleus is allowed to take
the values $Z_a=0,4$ and 8 (naturally this last
value is unphysical but intends to exaggerate the effect).

\begin{figure}[h]
\centerline{
        \parbox[t]{0.5\textwidth}{
\centerline{\psfig{figure=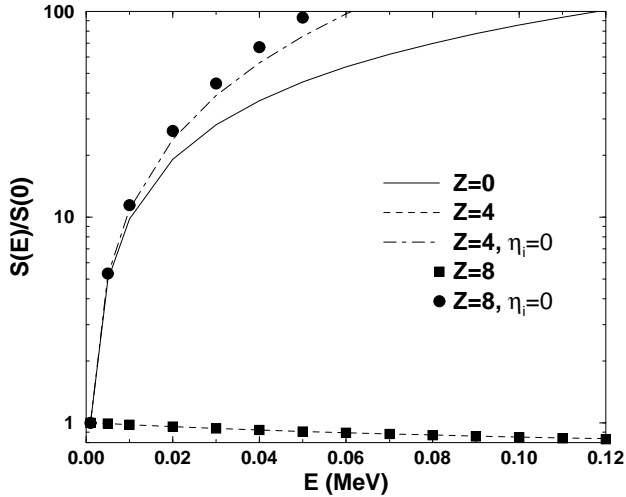,width=0.5\textwidth}}
\caption{S-factors for the E1-capture ${}^{7}{\rm Be}(p,
\gamma){}^{8}{\rm B}$: 
The solid line is the result for S(E) assuming $Z_{a}=0$;
the dashed line and dot-dashed correspond to S(E) for $Z_{a}=4$
calculated with and without the initial Coulomb barrier, correspondingly; 
the squares and circles correspond to S(E) for $Z_{a}=8$
calculated with and without the initial Coulomb barrier, correspondingly.
All S-factors are normalized to unity at zero energy.}
\label{fig_8Bli0}}
}
\end{figure}

\begin{figure}[htb]
\centerline{
        \parbox[t]{0.5\textwidth}{
\centerline{\psfig{figure=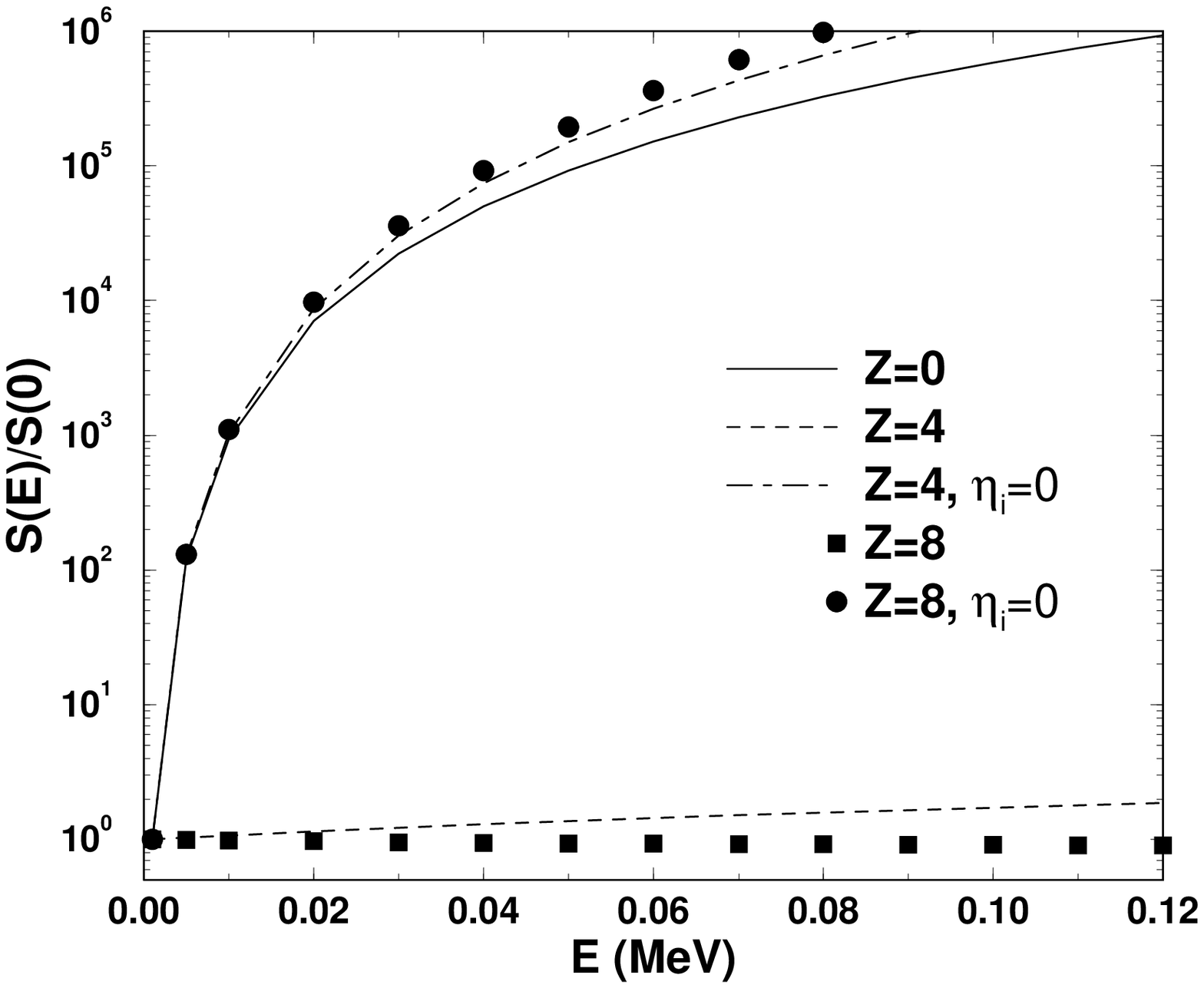,width=0.5\textwidth}}
\caption{S-factors for the same reaction as Fig.\ref{fig_8Bli0},
but  $l_{i}=2, \to l_{f}=1 $: 
The solid line is the result for S(E) assuming $Z_{a}=0$; 
the dashed line and dot-dashed correspond to S(E) for $Z_{a}=4$
calculated with and without the initial Coulomb barrier, correspondingly; 
the squares and circles correspond to S(E) for $Z_{a}=8$
calculated with and without the initial Coulomb barrier, correspondingly.
All S-factors are normalized to unity at zero energy.}
\label{fig_8Bli2}}
}
\end{figure}

In order to isolate these effects we run three different calculations.
First, the S-factors resulting from setting $Z_a=0$,
such that none of the Coulomb barrier is present.
Secondly, the results of the calculations
for which the remnant initial-state Coulomb barrier
is set to zero ($\eta_{i}=0$), but keeping $\eta_{f}$.
It contains thus only the final-state Coulomb barrier.
Finally, the full calculation for S(E), including both
initial and final Coulomb barriers.
Results are plotted
in Fig.(\ref{fig_8Bli0}) for a transition $1 \to 0$  and in 
Fig.(\ref{fig_8Bli2})
for a transition $2 \to 1$.
From these figures we conclude:
\begin{enumerate}
\item{the final Coulomb barrier increases the S-factor derivative,
as a normal penetration barrier would;}
\item{the larger the charge, the stronger the effect of the final-state
Coulomb barrier;}
\item{the effect of the final-state Coulomb barrier is significantly
weaker than the initial remnant Coulomb barrier;}
\item{for all cases, the initial remnant Coulomb barrier significantly
decreases the first derivative, acting as an attractive potential;}
\item{the larger the charge, the larger the decrease of $S'(0)$, such
that in most cases it wins over all other {\it real} barriers
making $S'(0)$ negative (note that this happens even for
$l_{i}=2$, where all previous cases had $S'(0)>0$).}
\end{enumerate}


\section{Comparative behavior of  {$S$}
factors for some important astrophysical processes}
\label{application}

\begin{figure}[htb]
\centerline{
        \parbox[t]{0.5\textwidth}{
\centerline{\psfig{figure=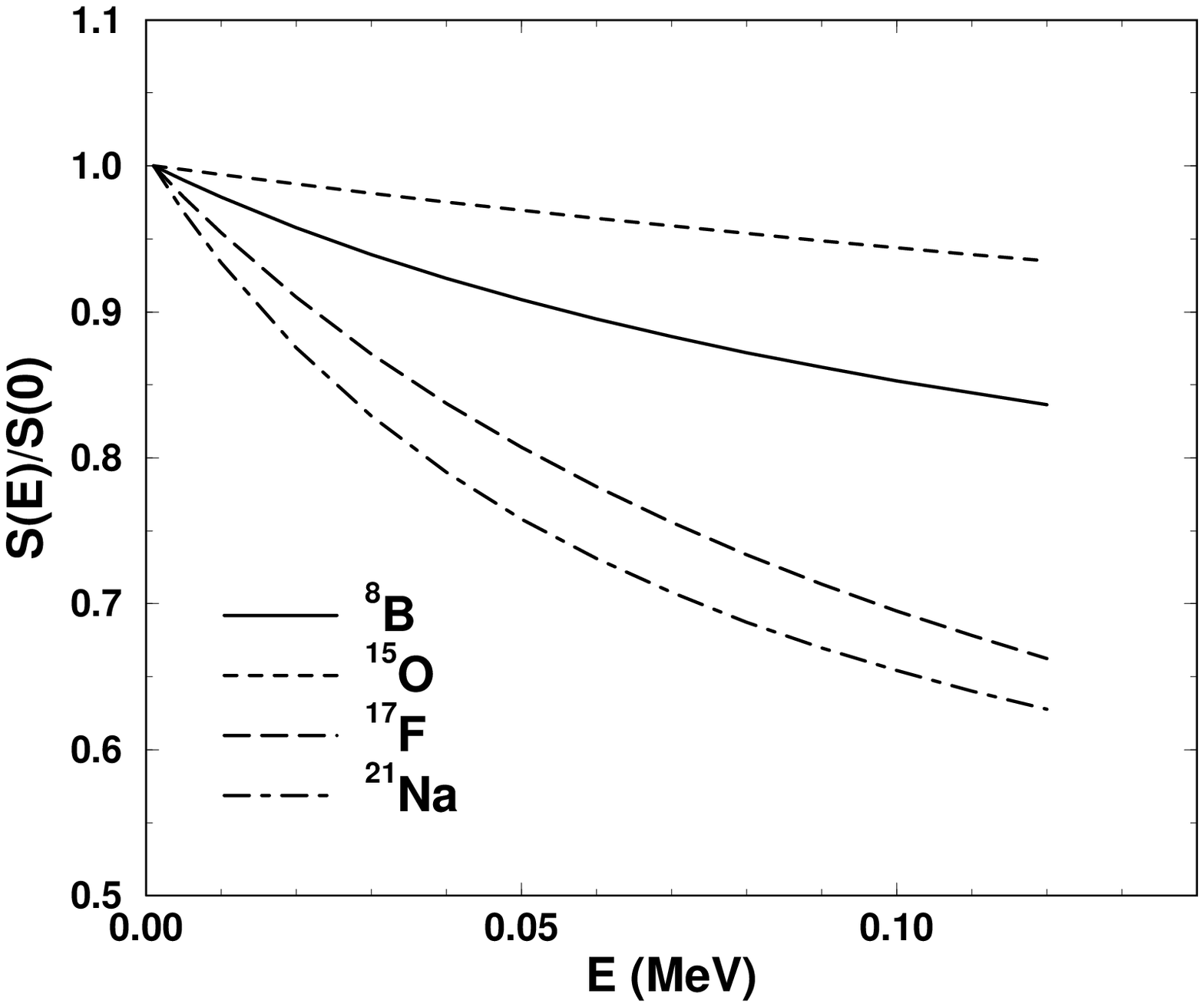,width=0.5\textwidth}}
        \caption{S-factors for some astrophysical reactions:
$^7{\rm Be}(p, \gamma)^8{\rm B}(2^{+})$, corresponding to a transition
$l_{i}=0 \to l_{f}=1$ with binding energy $\varepsilon=0.137$ MeV
(solid);
$^{14}{\rm N}(p, \gamma)^{15}{\rm O}(3/2^{+})$, corresponding to a
transition
$\;l_{i}=1 \to l_{f}=0$ with binding energy $\varepsilon=0.504$ MeV
(dotted);
$^{16}{\rm O}(p, \gamma)^{17}{\rm F}(1/2^{+})$, corresponding to a
transition
$l_{i}=1 \to l_{f}=0$ with binding energy $\varepsilon=0.105$ MeV
(dashed);
$^{20}{\rm Ne}(p,\gamma)^{21}{\rm Na}(1/2^{+})$, corresponding to a
transition
$l_{i}=1 \to l_{f}=0$ with binding energy $\varepsilon=0.0064$ MeV
(dot-dashed).
All the astrophysical factors are normalized to unity at zero energy.}
\label{fig_11}}
}
\end{figure}

In this section we compare and explain the difference in the energy
behavior of the direct capture S-factors for some important
astrophysical processes. Since we are only interested in the
comparative
energy behavior, all S-factors are normalized to unity at $E=0$.
Four notorious reactions with either $l_{i}=1$ or $l_{f} =1$
are considered: \\
a) ${}^{7}{\rm Be}(p, \gamma){}^{8}{\rm
B}(2^{+}),\;\,\varepsilon=0.137$ MeV,
$\;\,l_i=0 \to l_f=1$;\\
b) ${}^{14}{\rm N}(p, \gamma){}^{15}{\rm O}(3/2^{+}),
\;\;\varepsilon=0.504$ MeV,
$\;\,l_i=1 \to l_f=0$;\\
c) ${}^{16}{\rm O}(p, \gamma){}^{17}{\rm
F}(1/2^{+}),\;\;\varepsilon=0.105$ MeV,
$\;\,l_i=1 \to l_f=0$;\\
d) ${}^{20}{\rm Ne}(p, \gamma){}^{21}{\rm Na}(1/2^{+}),\;\;\varepsilon=
0.0064$
MeV, $\;\,l_i=1 \to l_f=0$.\\
Results are presented in Fig.\ref{fig_11}.
Since $l_{i},\,l_{f} \leq 1$, the centrifugal
barriers are not strong enough to win over the effect of the 
singularity at
$E=-\varepsilon$ and the initial remnant Coulomb barrier, leading
to negative slopes for $S(0)$.
Within these examples, the largest charge system,
also corresponds to the lowest binding energy, such that both effects
result in a very negative derivative for $S(0)$. A significant
difference
between $S'(0)$ for the $^8$B and $^{17}$F cases, with
similar binding energies, proves that the initial Coulomb remnant
barrier which is stronger for ${}^{17}$F than for ${}^{8}$B,
wins over the initial centrifugal barrier
$l_{i}=1$, resulting in a more negative derivative $S'(0)$
for ${}^{17}$F.  

\begin{figure}[htb]
\centerline{
        \parbox[t]{0.5\textwidth}{
\centerline{\psfig{figure=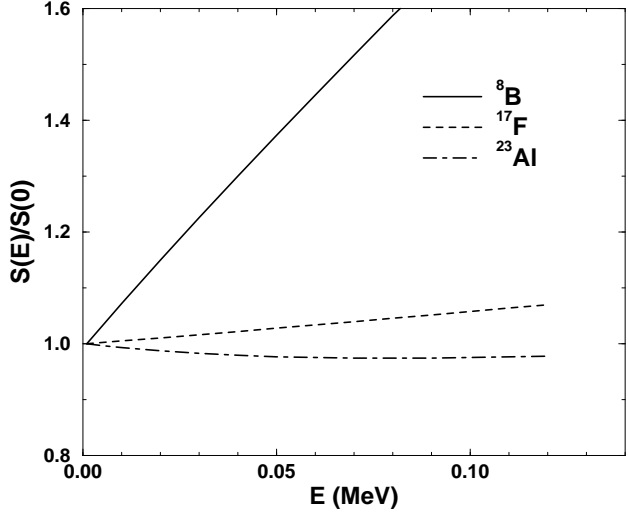,width=0.5\textwidth}}
        \caption{S-factors for some astrophysical reactions:
$^7$Be$(p, \gamma)^8$B$(2^{+})$, corresponding to a transition
$l_{i}=2 \to l_{f}=1$ and binding energy of the final state
$\varepsilon=0.137$ MeV (solid);
$^{16}$O$(p, \gamma)^{17}$F$(5/2^{+})$, corresponding to a transition
$l_{i}=1 \to l_{f}=2$, and binding energy of the final state
$\varepsilon=0.605$ MeV (dashed);
$^{22}$Mg$(p, \gamma)^{23}$Al$(5/2^{+})$, corresponding to a transition
$l_{i}=1 \to l_{f}=2$ and binding energy of the final state
$\varepsilon=0.127$ MeV (dot-dashed).
All the astrophysical factors are normalized to unity at zero energy.}
\label{fig_22}}
}
\end{figure}

We also consider astrophysical examples for which either
$l_{i}=2$ or $l_{f}=2$:\\
a) ${}^{7}{\rm Be}(p,\gamma){}^{8}{\rm B}(2^{+}),\;\varepsilon=0.137$
MeV, with $l_i=2 \to l_f=1$;\\
b) ${}^{16}{\rm O}(p,\gamma){}^{17}{\rm
F}(5/2^{+}),\;\,\varepsilon=0.605$ MeV,
with $l_i=1 \to l_f=2$;\\
c) ${}^{22}{\rm Mg}(p, \gamma){}^{23}{\rm
Al}(5/2^{+}),\,\,\varepsilon=0.127$ MeV,
with $l_i=2 \to l_f=1$.\\
The resulting S-factors are presented in Fig.\ref{fig_22}.
For the first two cases the slope is positive.
For the ${}^{8}$B case, the initial centrifugal barrier and smaller charge 
result in the highest $S'(0)$.
The fact that the remnant Coulomb barrier is stronger for $^{17}$F than
for $^8$B and that the initial centrifugal barrier is lower, 
produces a smaller $S'(0)$.
For the capture on $^{22}$Na, despite the high initial orbital
$l_{i}=2$, the very large initial remnant Coulomb barrier (together 
with
the trace of the nearby singularity) is able to make the slope
negative near zero.


\section{Summary}
\label{summary}

We investigated the low-energy behavior of the astrophysical $S(E)$
factors for direct radiative captures of protons by charged
nuclei leading to loosely bound final states.
For such processes a simple potential model,
extending the asymptotic form of the final state
to $r=0$ and assuming a pure Coulomb scattering wave in the initial
state are well justified, as most of the contribution to the
low energy capture to loosely bound states comes from large distances.

We demonstrate that the behavior of the S-factors is governed by
six essential ingredients.
Two act in an attractive sense, creating a negative slope for $S(0^+)$:
the remnant of the initial Coulomb barrier
(left after extracting the Gamow penetration factor) and the
singularity
at $E=-\varepsilon$ (where $\varepsilon$
is the binding energy of the final state).
Three act as {\em real} penetration barriers, producing
a positive slope for $S(0^+)$:
both initial and final centrifugal barriers and the final-state Coulomb
barrier. The effect of the final centrifugal and Coulomb barriers are
minor
compared with the initial centrifugal and Coulomb. 
Finally, there is still a photon factor $k_{\gamma}^{2\,L+1}$,
which tends to increase the derivative of $S(E)$.
We have derived analytical expressions for the S-factor
in a few typical cases.
We have tried to desmystified the idea that the energy behavior
of the S-factor around threshold is dominated by the pole
$E=-\varepsilon$.
We have also shown that by taking only the first term of the
Whittaker expansion (final state), one cannot reproduce the correct
S-factor
energy behavior at threshold, except when the Coulomb parameters
of the final state is integer.
Finally, we have not only illustrated our findings with a few sets
of study cases but applied it to specific examples relevant in
astrophysics.

Although in this work, we were mainly concerned with the energy 
behavior 
of the astrophysical factors, our approximate equations can reproduce, 
the absolute values of the capture rates within a few percent. 
The smaller the binding energy, the better accuracy
of our equations. The application of Eq. (\ref{sfctr2}) can be extended 
to energies of a few hundred keV, {\it i.e.} to the explosive 
nucleosynthesis 
energy range, when binding energies are not larger than 1 MeV. 
Specifically, for charges $Z_{a}=4 - 10$ and binding energies 
$\varepsilon \leq 1.0$ MeV,  S-factors obtained from Eq. (\ref{sfctr2}) 
do not deviate more than a few $\%$ from the exact potential model,
for relative energies up to 1 MeV.
However for incident energies $\geq 400$ keV 
and/or higher binding energies, a simple potential model becomes less 
accurate:
nuclear interactions in the initial state and microscopic effects,
such as antisymmetrization, should to be included.  
\vspace{0.5cm}

\acknowledgements
One of the authors, F.M. Nunes, wishes to thank M.A. Nagarajan for
many discussions at an earlier stage, which led to the derivation of
Eq. (\ref{sfint3}). Support from the Portuguese Science and Technology 
Foundation (FCT) under grant SAPIENS/36282/99 and the 
U.S. Department of Energy under grant DE-FG03-93ER40773 are 
acknowledged.
One of the authors, A.M. Mukhamedzhanov, was supported by
Centro Multidisciplinar de Astrofisica CENTRA, under
FCT plurianual 2/99.

\newpage
\setcounter{section}{1}
\setcounter{equation}{0}
\def\theequation{\Alph{section}.\arabic{equation}}
\begin{appendix}

\section{Integral representation of {$S(E)$}} \protect
\label{app1}
In the Appendix we derive the expression for  the S-factor for captures
to loosely bound states. Our starting point is the definition in
Eq. (\ref{sfctr2}).
We simplify this expression, reducing it to a first-order integral with
an integrand expressed in terms of elementary functions.
To this end, we use the integral representation for both bound and
scattering states, under the assumption of the asymptotic approximation
for the bound state wave function discussed in the main text. 

The Whittaker
function has the following integral representation \cite{gr80}:
\begin{eqnarray}
W_{-\eta_{f},l_{f}+1/2}(2\,\kappa\,r) =
\frac{(2\,\kappa\,r)^{l_{f}+1}\,e^{-\kappa\,r}}
{\Gamma(l_{f}+\eta_{f}+1)}\,
\int\limits_{0}^{\infty}\,{\rm d}s\,e^{-2\,\kappa\,r\,s}\,
\lbrack s(1+s) \rbrack^{l_{f}} \, \lbrack \frac{s}{1+s}
\rbrack^{\eta_{f}}.
\label{intwhit1}
\end{eqnarray}
Furthermore, the integral representation of the first term of its
asymptotic expansion in powers $1/\kappa\,r$ ($j=0$ in
Eq.\ref{asexpwh1}),
is given by:
\begin{eqnarray}
W_{-\eta_{f},l_{f}+1/2}^{(0)}(2\,\kappa\,r)&=&
\frac{e^{-\kappa\,r}}{(2\,\kappa\,r)^{\eta_{f}}}                           
\nonumber\\
& = &\frac{(2\,\kappa\,r)^{l_{f}+1}\,e^{-\kappa\,r}}
{\Gamma(l_{f}+\eta_{f}+1)}\,
\int\limits_{0}^{\infty}\,{\rm d}s\,e^{-2\,\kappa\,r\,s}\,
s^{l_{f}+\eta_{f}}.
\label{intwhitf0}
\end{eqnarray}
The difference between $W_{-\eta_{f},l_{f}+1/2}(2\,\kappa\,r)$ and the
first
term $W^{(0)}_{-\eta_{f},l_{f}+1/2}(2\,\kappa\,r)$ integral
representations
is only in the spectral function.

The Coulomb regular solution, $F_{l_{i}}(k,r)$, is
given by Eq. (\ref{eq:ss1}) in terms of the confluent hypergeometric
function. This can be written in integral form as \cite{gr80}:  
\begin{eqnarray}
{}_{1}F_{1}(l_{i}+1-i\,\eta_{i},\,2\,l_{i}+2,;\, 2\,i\,k\,r)&=&
\frac{1}{B(l_{i}+1-i\,\eta_{i},l_{i}+1+i\,\eta_{i})}\,   \nonumber\\
& \times & \int\limits_{0}^{1}\,{\rm d}t\,e^{2\,i\,k\,r\,t}\,
\lbrack t(1-t) \rbrack^{l_{i}}\, \lbrack \frac{1-t}{t}
\rbrack^{i\,\eta_{i}}.
\label{kumerint1}
\end{eqnarray}
Here, $B(l_{i}+1-i\,\eta_{i},l_{i}+1+i\,\eta_{i})$ is a
$\beta$-function
expressed in terms of $\Gamma$-functions:
\begin{equation}
B(l_{i}+1-i\,\eta_{i},l_{i}+1+i\,\eta_{i})=
\frac{\Gamma(2\,l_{i}+2)}
{\Gamma(l_{i}+1-i\,\eta_{i})\,\Gamma(l_{i}+1+i\,\eta_{i})}.
\label{bfnct1}
\end{equation}
Substituting Eq. (\ref{intwhit1}) and Eq. (\ref{kumerint1}) into
Eq. (\ref{sfctr2}),
allowing for Eq. (\ref{clet1}) and changing the order of the
integrations,
we arrive at the following expression for the S-factor:
\begin{eqnarray}
S(E)&=& {\cal A} \; 2^{2\,l_{i}+2\,l_{f}+4}\pi
\frac{1}{[\Gamma(l_{f}+\eta_{f}+1)]^{2}}\,
\frac{1}{[\Gamma(2\,l_{i}+2)]^{2}}               
\big|B(l_{i}+1-i\,\eta_{i},l_{i}+1+i\,\eta_{i})\big|^{2}
\nonumber\\
&\times &\, k_{\gamma}^{2\,L+1}\,\frac{\kappa^{2l_{f}+1}}{\eta_{f}}
\prod_{j=0}^{l_{i}}(j^{2}\,k^{2}+ \eta_{f}^{2}\,\kappa^{2})   \; |{\cal
J}|^2,
\label{sfctra1}
\end{eqnarray}
We have merged the three integrations into ${\cal J}$:
\begin{eqnarray}
{\cal J} & = & \int\limits_{0}^{\infty}{\rm d}s\,\lbrack s(1+s)
\rbrack^{l_{f}}
\lbrack \frac{s}{1+s} \rbrack^{\eta_{f}}\,
\int\limits_{0}^{1}{\rm d}t \lbrack t(1-t)\rbrack^{l_{i}} \lbrack
\frac{1-t}{t}
\rbrack^{i\eta_{i}}
\nonumber \\
& \times &
\int\limits_{r{0}}^{\infty}{\rm d}r\,r^{L+l_{i}+l_{f}+2}
e^{-r\lbrack \kappa\,(1+2\,s) + i\,k\,(1-2\,t)\rbrack }\;.
\end{eqnarray} 
We now replace the lower limit of the integration by $r_0=0$,
keeping in mind the application to loosely bound states (where the
contribution from the interior is not significant).
The result, after integrating over $r$, holds:
\begin{eqnarray}
{\cal J} &=&  {\cal V}\, \int\limits_{0}^{\infty}\,{\rm d}s\,
\overbrace{\lbrack s(1+s) \rbrack^{l_{f}} \, \lbrack \frac{s}{1+s}
\rbrack^{\eta_{f}}}^{{\cal S}(s)} \,  \int\limits_{0}^{1}\,{\rm d}t\,
\lbrack t(1-t) \rbrack^{l_{i}}\, \lbrack \frac{1-t}{t}
\rbrack^{i\,\eta_{i}}
\nonumber\\
& \times & \,\frac{1}
{\lbrack \kappa\,(1+2\,s) + i\,k\,(1-2\,t)\rbrack^{l_{i}+l_{f}+L+3}},
\label{int1}
\end{eqnarray}
with ${\cal V}= \int\limits_{0}^{\infty}\,{\rm
d}v\,v^{l_{i}+l_{f}+L+3}\,e^{-v}$.
Let us define the spectral function of $s$, as indicated
above, ${\cal S}(s)$ and the remaining integral over $t$ as ${\cal
J}_t$.
The expression for ${\cal S}(s)$ is:
\begin{equation}
{\cal S}(s) = \lbrack s(1+s) \rbrack^{l_{f}} \, \lbrack \frac{s}{1+s}
\rbrack^{\eta_{f}}\,.
\label{intSs}
\end{equation}
The integral over $t$ can be reduced to:
\begin{eqnarray}
{\cal J}_{t} & \equiv &
\int\limits_{0}^{1}\,{\rm d}t\,
\lbrack t(1-t) \rbrack^{l_{i}}\, \lbrack \frac{1-t}{t}
\rbrack^{i\,\eta_{i}}\,\frac{1}
{\lbrack \kappa\,(1+2\,s) + i\,k\,(1-2\,t)\rbrack^{l_{i}+l_{f}+L+3}}
\nonumber\\
& = & (-1)^{m}\,\frac{(2\,l_{i}+1)!}{(l_{f}+L+l_{i}+2)!}
\nonumber \\
& \times & \lim_{\epsilon \to 0}\,\frac{\rm d^{m}}
{\rm d\, \epsilon^{m}}\, \int\limits_{0}^{1}\,{\rm d}t\,
\lbrack t(1-t) \rbrack^{l_{i}}\, \lbrack \frac{1-t}{t}
\rbrack^{i\,\eta_{i}}\, \frac{1}{\lbrack \kappa\,(1+2\,s)+ \epsilon +
i\,k\,(1-2\,t)\rbrack^{2\,l_{i}+2}} \;,
\label{intt2}
\end{eqnarray}
where we introduce $m=l_{f}+L+1-l_{i}$.
One can further simplify the integrand by factorizing the t-independent
terms, such that the remaining integral
consists only on the integral representation of the hypergeometrical
function $_2F_1$:
\begin{eqnarray}
{\cal J}_{t} & = & (-1)^{m}\,\frac{(2\,l_{i}+1)!}{(l_{f}+L+l_{i}+2)!}\,
B(l_{i}+1-i\,\eta_{i},l_{i}+1+i\,\eta_{i})                  
\nonumber\\
& \times & \lim_{\epsilon \to 0}\,\frac{\rm d^{m}}{\rm d
\epsilon^{m}}\,
\frac{1}{\lbrack \kappa\,(1+2\,s)+ \epsilon + 
i\,k\rbrack^{2\,l_{i}+2}}     
{}_{2}F_{1}(l_{i}+1-i\,\eta_{i}, 2\,l_{i}+ 2,2\,l_{i}+2; y).
\label{intt3}
\end{eqnarray}
The argument of the hypergeometric Gauss function is
$y=1 - \frac{2\,i\,k}{\kappa\,(1+2\,s)+ \epsilon +
i\,k} $. Taking into account the property:
${}_{2}F_{1}(\alpha, \beta,\gamma; y) = (1- y)^{\gamma - \alpha -
\beta}\,
{}_{2}F_{1}(\gamma - \alpha, \gamma - \beta,\gamma; y)$
(Eq.9.131-1 of \cite{gr80}), and the fact that
$_{2}F_{1}(\alpha, 0,\gamma; y)=1$,
we can arrive at a simplified expression:
\begin{eqnarray}
{\cal J}_{t} &=& (-1)^{m}\,\frac{(2\,l_{i}+1)!}{(l_{f}+L+l_{i}+2)!}\,
B(l_{i}+1-i\,\eta_{i},l_{i}+1+i\,\eta_{i})
\nonumber\\
& \times &\lim_{\epsilon \to 0}\frac{\rm d^{m}}{\rm d\,\epsilon^{m}}             
\frac{1}{\lbrack (\kappa\,(1+2\,s)+ \epsilon)^{2} +
k^{2}\rbrack^{l_{i}+1}} 
e^{2 \eta_{i}\arctan \frac{k}{\kappa(1+2\,s)+ \epsilon}}.
\label{intt4}
\end{eqnarray}

Introducing Eq. (\ref{intt4}) and Eq. (\ref{intSs}) into Eq. 
(\ref{int1}),
${\cal J}$ reduces to
\begin{eqnarray}
{\cal J} &=&  {\cal V} \, (-1)^{m}\,
\frac{(2\,l_{i}+1)!}{(l_{f}+L+l_{i}+2)!}\,
B(l_{i}+1-i\,\eta_{i},l_{i}+1+i\,\eta_{i}) \,
\nonumber\\
&&\times \int\limits_{0}^{\infty}{\rm d}s\, {\cal S}(s) \,
\lim_{\epsilon \to 0}\,
\frac{\rm d^{m}}{\rm d\, \epsilon^{m}}\,
\frac{1}{\lbrack (\kappa\,(1+2\,s)+
\epsilon)^{2}+k^{2}\rbrack^{l_{i}+1}}\,
e^{2 \eta_{i}\arctan \frac{k}{\kappa(1+2\,s)+ \epsilon}}\,.
\label{int2}
\end{eqnarray}
Consequently, replacing Eq. ({\ref{int2}) in Eq. (\ref{sfctra1}),
$S(E)$ takes the form:
\begin{eqnarray}
S(E)&=& {\cal B} \;
k_{\gamma}^{2\,L+1}\,\frac{\kappa^{2l_{f}+1}}{\eta_{f}}\,
\prod_{j=0}^{l_{i}}(j^{2}\,k^{2}+ \eta_{f}^{2}\,\kappa^{2})
\nonumber\\
&&\times \bigg|\lim_{\epsilon \to 0}\,
\int\limits_{0}^{\infty}{\rm d}s\, {\cal S}(s) \,
\frac{\rm d^{m}}{\rm d\, \epsilon^{m}}\,
\frac{1}{\lbrack (\kappa\,(1+2\,s)+
\epsilon)^{2}+k^{2}\rbrack^{l_{i}+1}}\,
e^{2\,\eta_{i}\arctan \frac{k}{\kappa(1+2\,s)+ \epsilon}}\bigg|^{2},
\label{sfctrf5}
\end{eqnarray}
where a new constant, putting in evidence energy independent factors,
is introduced:
\begin{equation}
{\cal B} = {\cal A} \; 2^{2\,l_{i}+2\,l_{f}+4}\,\pi
\frac{1}{[\Gamma(l_{f}+\eta_{f}+1)]^{2}}\,  
\frac{[(2\,l_{i}+1)!]^{2}}{[(l_{f}+L+l_{i}+2)!]^{2}}\,
\frac{1}{[\Gamma(2\,l_{i}+2)]^{2}}   \,.
\label{constB}
\end{equation}
The constant ${\cal A}$ was introduced in the beginning of the
text
by Eq. (\ref{eq:constant}).

Similarly, the function $S_{(0)}(E)$, resulting from the first term
of the Whittaker expansion, can be derived. The result holds an
identical
expression to $S(E)$, the only difference being the spectral function
over $s$.
Thus, Eq. (\ref{sfctrf5}) is applicable to $S_{(0)}(E)$ when replacing
${\cal S}(s)$ by ${\cal S}_{0}(s)$:
\begin{equation}
{\cal S}_{0}(s) =  s^{l_f+\eta_f}.
\label{intSs0}
\end{equation}

It is important to note that the replacement of the lower
integration limit in the radial integral of Eq. (\ref{sfctra1})
imposes conditions for the convergence
of the integrals over $s$  at $\infty$: \\
1) for $S(E)$, $l_{f} < 2 + L + l_{i}$, which is always satisfied;\\
2) for $S_{(0)}(E)$, $\eta_{f} < 2 + L + l_{i}$, which limits the
charge
of the capturing nucleus.

Finally, we should mention that, as we focus only on dipole
transitions,
then $|l_{i}-l_{f}|=1$ and $L=1$. Hence, there
are two possibilities for $m$: $l_{i} - l_{f} = 1 \rightarrow m=1$  and
$l_{f}-l_{i}= 1 \rightarrow m=3$.  In each case an evaluation of
Eq. (\ref{sfctrf5}}) can be explicitly performed.
This is presented and discussed in section \ref{thresh}.

\end{appendix}

\vspace{-0.5cm}

\end{document}